\documentclass[pra,twocolumn,superscriptaddress,showpacs,floatfix]{revtex4}

\usepackage{graphics,epsf,times,amsfonts,bm,bbm}
\usepackage{latexsym}
\usepackage{amsthm}
\usepackage{amssymb}
\usepackage{amsmath}
\usepackage{graphicx}
\usepackage[
colorlinks=true,linkcolor=red,citecolor=green,urlcolor=blue,
pdfstartview=FitH,
bookmarksopen=true,bookmarksopenlevel=0,
pagebackref=false,
pdfauthor={Soraya Taghavi, Todd A. Brun, Daniel A. Lidar}, 
pdftitle={Optimized Entanglement-Assisted Quantum Error Correction}
pdfsubject={quantum information processing}, 
pdfkeywords={quantum, information, computation, error correction,
  entanglement, optimization}]{hyperref}
\usepackage{hypernat}

\newcommand{\ignore}[1]{}
\include{qec_texdefs}

\begin{document}

\title[]{Optimized Entanglement-Assisted Quantum Error Correction}
\author{Soraya Taghavi}
\affiliation{Department of Electrical Engineering, Center for
Quantum Information Science \& Technology, University of Southern California, Los
Angeles, California 90089, USA}
\author{Todd A. Brun}
\affiliation{Department of Electrical Engineering, Center for
Quantum Information Science \& Technology, University of Southern California, Los
Angeles, California 90089, USA}
\author{Daniel A. Lidar}
\affiliation{Departments of Electrical Engineering, Chemistry, and Physics, Center for
Quantum Information Science \& Technology, University of Southern California, Los
Angeles, California 90089, USA}


\begin{abstract}
Using convex optimization, we propose entanglement-assisted quantum error
correction procedures that are optimized for given noise channels. We
demonstrate through numerical examples that such an optimized error
correction method achieves higher channel fidelities than existing methods.
This improved performance, which leads to perfect error correction for a
larger class of error channels, is interpreted in at least some cases by
quantum teleportation, but for general channels this interpretation does not
hold.
\end{abstract}

\date{\today}
\maketitle

\section{Introduction}

Noise is an important obstacle for the scale-up of quantum information
devices---both decoherence due to interactions with an external environment,
and the internal control errors inevitable for any information-processing.
The theory of quantum error correction was developed by analogy to classical
error-correcting codes to overcome this obstacle \cite{Shor:95,Steane:96a,Gottesman:96,Knill:97b}. Quantum error-correcting codes
(QECCs) store quantum information redundantly in an entangled state of
multiple quantum systems (usually qubits), in such a way that errors can be
detected and corrected without directly measuring (and hence disturbing) the
quantum information to be protected. The fact that this can be done at all
is quite remarkable, and has opened up the prospect that quantum
information processing may become a viable technology in the foreseeable future.

Broadly speaking, quantum error correction is useful in two main contexts.
The first is quantum computation, where errors that occur during a
calculation must be constantly found and corrected to keep them from
accumulating. The QECCs used for this purpose must fit into an overall
fault-tolerant scheme \cite{Aharonov:96,Kitaev:96,Gottesman:97a,Preskill:97a}%
, which puts important constraints on the properties of the codes used.

The other main context is quantum communication. Here it is assumed that the
steps of encoding and decoding are largely error-free, and all errors occurs
during the transmission of quantum information through a noisy quantum
channel. This is the context we will consider in this paper. It should be
noted, though, that this division is not really sharp: quantum computers may
well need to transmit information internally, and quantum communication may
be part of a distributed information processing scheme.

In the context of quantum communications, one can explore the effect of
different shared resources on the ability to detect and correct errors.
Recently, a great deal of work has been done on entanglement-assisted
quantum error correction (EAQEC), and entanglement-assisted quantum
error-correcting codes (EAQECCs). In these schemes, the sender (usually
called Alice) and receiver (usually called Bob) share some number of pure
maximally-entangled states, or ebits. Alice can use her halves of these
ebits in the encoding procedure, and Bob can use his halves in error
correction and decoding. The idea of using entanglement to improve
error correction was proposed early in the history of quantum information
theory \cite{Breeding:96}, but it took a long time for the first steps towards
designing practical codes \cite{Bowen:02}.  This work has led to a
generalized theory of quantum error correcting codes, the
entanglement-assisted stabilizer formalism
\cite{Brun:06,Brun:06b,Hsieh:07,Wilde:08,Kremsky:08,Hsieh:09,Wilde:10}.

The large majority of work on QECCs has concerned stabilizer codes
\cite{Gottesman:96}, which can be derived from classical linear codes. In this 
algebraic approach, one generally tries to design codes that are able to
detect and correct an abstract set of errors acting on single qubits, often
based on the Pauli operators. QECCs have the property that any linear
combination of correctable errors is also a correctable error. Since the
Pauli operators form a basis for all single-qubit operators, the ability to
correct Pauli-based errors implies the ability to correct more realistic
models of noise, provided that errors are not too highly correlated between
qubits. The EAQECCs that have mostly been derived so far extend this
stabilizer formalism to include a broader class of codes that utilize shared
entanglement.

The standard approach to QEC finds encoding and recovery procedures
which ensure perfect recovery of quantum states
passing through sufficiently weak noisy channels. The disadvantage of
this approach is that it may fail when the noise strength, or error probability, is too
high. Another approach to QEC is to tailor QECCs
to a particular model of realistic noise, derived from a detailed
model or from experimental
measurements \cite{PhysRevLett.90.193601,Weinstein:04,Howard:06}, e.g., via quantum process tomography
\cite{Chuang:97c,Poyatos:97,DAriano:01,MohseniLidar:06,EmersonScience:07,MohseniRezakhaniLidar,Branderhorst:09}.
The tailored approach to QEC attempts to find encoding
and recovery operations which are \emph{optimized} relative to the particular
experimentally measured noise or assumed noise model, in the sense
that they maximize the fidelity (or minimize the distance) of the
encoded output state relative to the input state 
\cite{ReimpellW:05,YamamotoHT:05,FletcherSW:06,Fletcher:08,PhysRevA.77.012320,KosutLidar:06,KosutSL:07,TaghaviKL:qec07,PhysRevA.80.012326,PhysRevLett.104.120501,PhysRevA.81.062342,Tyson:09}.
Both worst case and average case performance has been considered in
this setting, which usually involves numerical optimization. The
advantage of this approach is that it tends to be more robust to variations in
the noise strength than the standard approach
\cite{TaghaviKL:qec07}.
Here we further develop the optimized QEC approach by considering, for the first time,
entanglement-assisted quantum error correction as an optimization
problem.

To formulate the problem for optimization, we break down quantum error
correction into three stages. First, the quantum state to be protected is
encoded. This procedure appends some number of ancilla qubits in a fixed
initial state, and then applies an encoding unitary to the information and
ancilla qubits together. These qubits then pass through a noisy channel
(assumed to be known). At the receiver's side, more ancillas can be
appended, and then a decoding unitary is applied. All the ancillas (encoding
or recovery) are discarded, and the state of the information qubits is the
output.

In standard quantum error correction, the encoding and recovery ancillas
always start in a standard state (usually $|0\rangle$), and are not
entangled with each other. Previous work has shown that regular recovery
ancillas are redundant in increasing the error correction fidelity \cite{TaghaviKL:qec07}. That is, given the same number of information qubits and
encoding ancillas, the presence or absence of recovery ancillas makes no
difference to the channel fidelity. This matches the properties of
stabilizer codes, where decoding can be done unitarily, and there is no
benefit in including recovery ancillas.

To formulate the problem for entanglement-assisted codes, we change the
initial state of the ancillas, so that noiseless entanglement is shared
between the encoding and recovery ancilla qubits. We show that this
additional information can enhance the performance of the recovery ancillas,
and increase the error correction fidelity in most cases. This initial
shared entanglement can in principle improve performance in at least two
different ways. It can allow the code to transfer quantum information about
the initial state to the recovery block, improving the channel fidelity.
This fidelity increase can in fact lead to perfect error correction for an
important class of error channels. The error correction procedure for this
class of channels can be interpreted as teleportation, with only classical
information passing through the noisy channel.

However, even in cases where the error correction protocol is not a form of
teleportation, we can still get an improvement in performance. One
interpretation of this is that entangled ancillas allow the recovery
procedure to extract more information about the errors. This is analogous to
superdense coding, where the use of entanglement boosts the rate of
classical communication. The fidelity increase here shows the importance of
relevant information stored in the ancillas in quantum error correction. The
entangled recovery ancillas are fundamentally different than the regular
recovery ancillas considered in previous optimization problems, that gave no
benefit \cite{TaghaviKL:qec07}.

The optimization procedure here, similar to previous work in this area \cite{ReimpellW:05,YamamotoHT:05,FletcherSW:06,Fletcher:08,PhysRevA.77.012320,KosutLidar:06,KosutSL:07,TaghaviKL:qec07,PhysRevA.80.012326,PhysRevLett.104.120501,PhysRevA.81.062342,Tyson:09},
assumes that we know the noise channel. We assume that a channel 
identification procedure, such as quantum process tomography \cite{PhysRevLett.90.193601,Weinstein:04,Howard:06,Chuang:97c,Poyatos:97,DAriano:01,MohseniLidar:06,EmersonScience:07,MohseniRezakhaniLidar,Branderhorst:09}, has been
performed prior 
to the error correction. It is also possible to optimize performance in
cases where the exact channel model is not known, though we do not consider
that here.

The structure of the paper is as follows. In Section~\ref{problem} we
formulate the optimization problem in terms of a distance between
states that needs to be minimized. The problem is bi-convex in the
encoding and recovery operations, and in Sections~\ref{sec:encoding} and
\ref{recovery} we consider how to optimize the encoding for a given
recovery, and recovery for a given encoding, respectively. In
Section~\ref{RUC} we consider the case of random unitary channels, and
analyze a simple example for which EAQEC has perfect fidelity, via
teleportation. We present numerical examples in
Section~\ref{examples}, where we contrast different correction and
optimization scenarios. These
examples illustrate the improved performance of optimized EAQECCs in a
setting where perfect fidelity cannot be achieved. We
conclude in Section~\ref{conc}. Certain technical details are
presented in Appendix~\ref{d5-proof}.

\section{Problem Formulation}
\label{problem}

A quantum dynamical process is a map from an initial state $\rho $ to a
final state $\mathcal{E}(\rho )$. In general the map $\mathcal{E}$ does not
represent unitary evolution. However, if we assume a larger closed quantum system that includes
the environment (or bath) $B$ of the system, the total evolution becomes
unitary. If the initial system-bath state is $\rho _{SB}$, a quantum channel
can be written as 
\begin{equation}
\mathcal{E}(\rho )=\mathrm{Tr}_{B}[U_{SB}\rho _{SB}U_{SB}^{\dagger }],
\label{channelU}
\end{equation}%
where $\rho =\mathrm{Tr}_{B}[\rho _{SB}]$ is the initial system density
matrix, $U_{SB}$ is a unitary operator acting on the joint system-bath
Hilbert space $\mathcal{H}=\mathcal{H}_{S}\otimes \mathcal{H}_{B}$, $\chi
_{B}=\mathrm{Tr}_{S}[\rho _{SB}]$ is the initial state of the environment,
and $\mathrm{Tr}_{B}$ ($\mathrm{Tr}_{S}$) denotes the partial trace over the
environment (system). In general a quantum dynamical process described by
Eq.~(\ref{channelU}) is a Hermitian map (i.e., it maps Hermitian operators
to Hermitian operators) \cite{ShabaniLidar:08}. When the initial system-bath
state is purely classicaly correlated (formally, has vanishing quantum
discord \cite{Ollivier:01}), $\mathcal{E}$ becomes a completely positive
(CP), trace-preserving map \cite{Rodriguez:07}. This condition is also
necessary \cite{ShabaniLidar:08}. In this case the evolution can be
represented using the Kraus operator sum representation (OSR) \cite{Kraus:83,Nielsen:book,Breuer:book}:
\begin{equation}
\mathcal{E}(\rho )=\sum_{i}K_{i}\rho K_{i}^{\dagger },
\end{equation}%
where the operators $K_{i}$, known as Kraus operators, satisfy the
normalization condition $\sum_{i}K_{i}^{\dagger }K_{i}=I$ (identity). Such
maps are powerful tools which can represent different types of noise
processes that could be acting on the system.

\begin{figure}[bp]
\includegraphics[width=8.5cm]{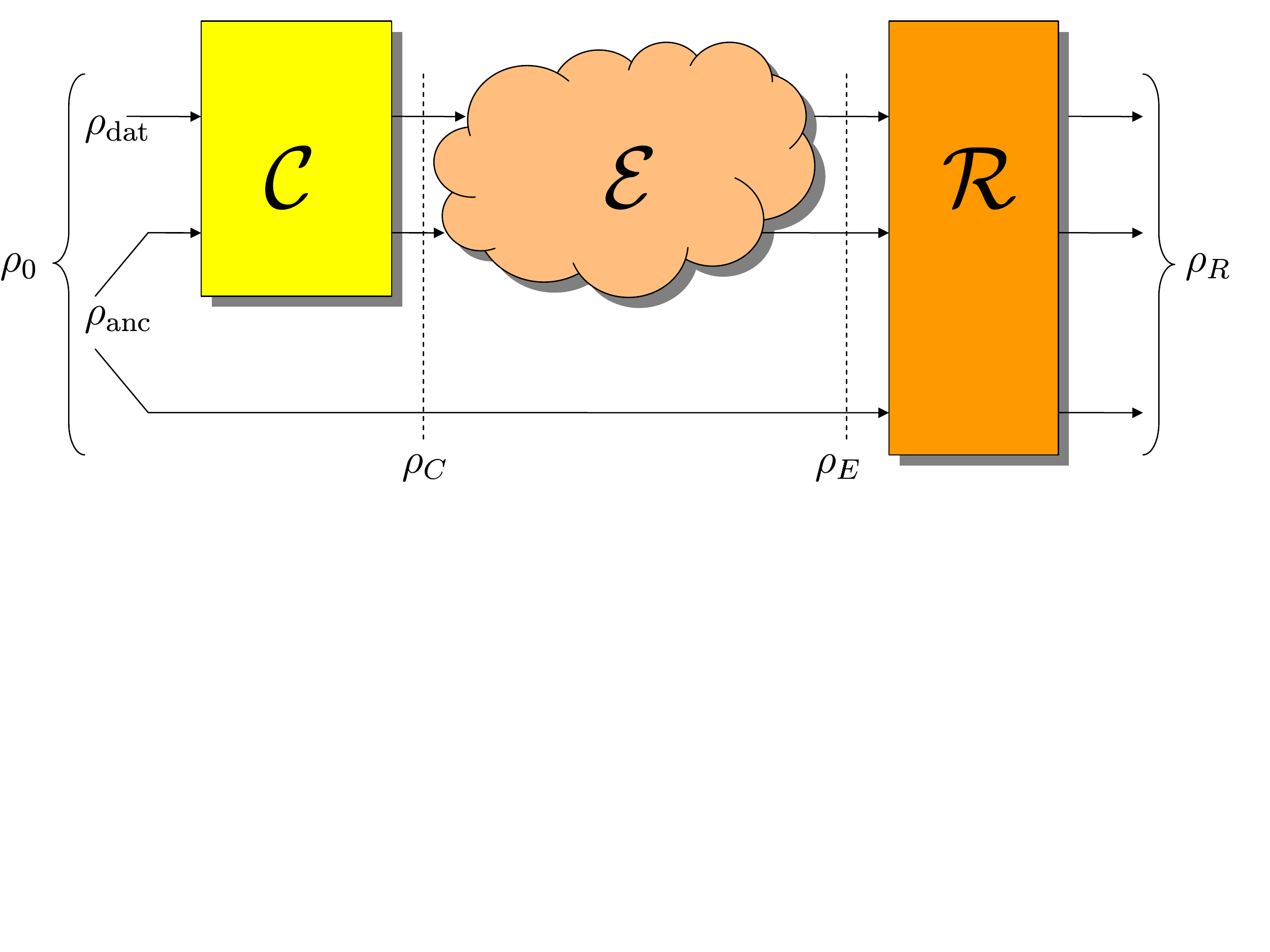} \vspace{-3.cm}
\caption{(Color online) Scheme of entanglement-assisted error correction.
The data qubits are represented by $\protect\rho _{\rm dat}$. The encoding
and recovery ancillas are entangled before the encoding procedure $\mathcal{C%
}$ begins, and are represented by $\protect\rho _{\rm anc}$. The joint initial
system state is $\protect\rho _{0}=\protect\rho _{\rm dat}\otimes
\protect\rho_{\rm anc} $. The recovery ancillas pass through a noiseless channel to the
recovery operation $\mathcal{R}$. The information qubits and encoding
ancillas pass through a noisy channel $\mathcal{E}$. The output of the
recovery procedure is $\protect\rho _{R}$. The intermediate states $\protect%
\rho _{C}$ (output of the encoding) and $\protect\rho _{E}$ (output of the
noise channel) are also indicated.}
\label{ea}
\end{figure}

We shall assume that the error correction procedures of encoding and
recovery, as well as the noisy channel, can all be described using Kraus
operators. The entire procedure on the system can then be represented as 
\begin{equation}
\rho _{0}\overset{\mathcal{C}}{\rightarrow }\rho _{C}\overset{\mathcal{E}}{%
\rightarrow }\rho _{E}\overset{\mathcal{R}}{\rightarrow }\rho _{R},
\end{equation}%
where $\mathcal{C}$, $\mathcal{E}$, and $\mathcal{R}$ are the encoding,
error and recovery operations respectively. The density matrices $\rho _{0}$
through $\rho _{R}$ are the states of the data qubits and the ancillas,
i.e., represent the state of the entire system at different points in time.
The initial system state $\rho _{0}:\mathcal{H}_{S}\mapsto \mathcal{H}_{S}$,
where $\mathcal{H}_{S}$ is the total system Hilbert space, is taken to be a
product state between the data qubits, represented by $\rho _{\mathrm{dat}}$%
, and the ancillas, represented by $\rho _{\mathrm{anc}}$, i.e.,\ $\rho
_{0}=\rho _{\mathrm{dat}}\otimes \rho _{\mathrm{anc}}$. Thus $\mathcal{H}%
_{S}=\mathcal{H}_{\mathrm{dat}}\otimes \mathcal{H}_{\mathrm{anc}}$, where $%
\mathcal{H}_{\mathrm{dat}}$ and $\mathcal{H}_{\mathrm{anc}}$ are,
respectively, the Hilbert space of the data qubits and of the ancillas. The
ancilla Hilbert space further decomposes into an encoding subspace $\mathcal{%
H}_{\mathrm{enc}}$ and a recovery subspace $\mathcal{H}_{\mathrm{rec}}$: $%
\mathcal{H}_{\mathrm{anc}}=\mathcal{H}_{\mathrm{enc}}\otimes \mathcal{H}_{%
\mathrm{rec}}$.\ We write $d=\dim (\mathcal{H}_{S})$ and otherwise denote
the respective dimensions of the various subspaces by: 
\begin{equation}
d_{X}=\dim (\mathcal{H}_{X}),\quad X\in \{\mathrm{dat},\mathrm{anc},\mathrm{%
enc},\mathrm{rec}\}.
\end{equation}%
Note that%
\begin{equation}
d=d_{\mathrm{dat}}d_{\mathrm{anc}},\quad d_{\mathrm{anc}}=d_{\mathrm{enc}}d_{%
\mathrm{rec}}.
\end{equation}

For most of our discussion we shall assume that all encoding ancillas are
pairwise maximally entangled with recovery ancillas prior to the encoding
operation, whence $d_{\mathrm{enc}}=d_{\mathrm{rec}}$, and $\rho _{\mathrm{%
anc}}=(\left\vert B\right\rangle \left\langle B\right\vert )^{\otimes d_{%
\mathrm{enc}}}$, where $\left\vert B\right\rangle $ is a maximally entangled
state between encoding and recovery ancilla pairs. Nevertheless, for
generality we shall leave $d_{\mathrm{enc}}$ and $d_{\mathrm{rec}}$ unless
it is necessary to be more specific. Later, in our numerical examples, we
shall also consider regular (unentangled) encoding and recovery ancillas.

Using the OSR for each of these operations, the overall evolution is 
\begin{eqnarray}
\rho _{R} &=&\mathcal{R}\circ \mathcal{E}\circ \mathcal{C}(\rho _{0})  \notag
\\
&=&\sum_{c,e,r}(R_{r}E_{e}C_{c})\rho _{0}(R_{r}E_{e}C_{c})^{\dagger },
\label{map}
\end{eqnarray}%
where $\{C_{c}\}_{c=1}^{m_{C}}$, $\{E_{e}\}_{e=1}^{m_{E}}$, and $%
\{R_{r}\}_{r=1}^{m_{R}}$ are the Kraus operators of the encoding, error, and
recovery channels, respectively, all satisfying the normalization condition: 
\begin{equation}
\sum_{c}C_{c}^{\dagger }C_{c}=\sum_{e}E_{e}^{\dagger
}E_{e}=\sum_{r}R_{r}^{\dagger }R_{r}=I.  \label{norm}
\end{equation}%
All the Kraus operators are here represented by square $d\times d$ matrices,
i.e., are operators on $\mathcal{H}_S$. The number of Kraus operators for each
map depends on the implementation and the basis representation, but the
number $m_{E}$ of error operators is bounded above by $d^{2}$ \cite{Nielsen:book}.

We shall assume that the encoding operation $\mathcal{C}$ is unitary,
represented by a single unitary (Kraus) operator $C$. The encoded $d$%
-dimensional state is therefore%
\begin{equation}
\rho _{C}=C(\rho _{\mathrm{dat}}\otimes \rho _{\mathrm{anc}})C^{\dagger }.
\end{equation}%
Since, as illustrated in Figure \ref{ea}, the encoding does not act on the
second half of the entangled pairs (the recovery ancillas), the encoding
unitary $C$ should be represented as $C^{\prime }\otimes I_{\mathrm{rec}}$,
where $C^{\prime }$ is a $d_{C^{\prime }}\times d_{C^{\prime }}$ matrix, and
where $d_{C^{\prime }}=d_{\mathrm{dat}}d_{\mathrm{enc}}=d_{\mathrm{dat}}%
\sqrt{d_{\mathrm{anc}}}$. We shall write $I_{X}$ to denote the identity
operator acting on a subspace $X$, and reserve $I$ for the identity operator
on the entire system space $\mathcal{H}_S$.

Our goal is to implement a close approximation to a certain desired unitary $%
L$ on the entire system space. More specifically, the purpose of
optimization is to design the encoding $\mathcal{C}$ and recovery $\mathcal{R%
}$ for a given error channel $\mathcal{E}$ such that the overall map (\ref%
{map}) $\rho _{0}\mapsto \rho _{R}$ is as close as possible to some desired
unitary $L:\mathcal{H}_{S}\mapsto \mathcal{H}_{S}$. This goal is different
from that pursued in earlier optimization papers (e.g., \cite{KosutLidar:06,KosutSL:07,TaghaviKL:qec07}), where the goal was to obtain a
good approximation to a unitary $L_{\mathrm{dat}}:\mathcal{H}_{\mathrm{dat}%
}\mapsto \mathcal{H}_{\mathrm{dat}}$ acting only on the data qubits. Our
present goal is general enough to encapsulate state preservation of the data
qubits, e.g., by implementing, with as high a fidelity as possible, a
unitary which swaps the data qubits into the recovery qubits at the end of
the process, while leaving the data and encoding qubits in a new pure state
which is tensored with the recovery qubits. Indeed, this is essentially what
happens in our teleportation example (Section \ref{RUC}).

Channel fidelity or distance are typical measures of performance between two
quantum channels \cite{Nielsen:book,GilchristLN:04,Kretschmann:04,ReimpellW:05}. The channel
fidelity between the error correction operation $\mathcal{REC}$ and the
desired unitary operation $L$ is: 
\begin{equation}
f=\frac{1}{d_{\mathrm{dat}}^{2}}\sum_{r,e,}\left\vert \mathrm{Tr}\
L^{\dagger }R_{r}E_{e}C\right\vert ^{2},  \label{Fidelity}
\end{equation}%
We wish to maximize this fidelity, which satisfies $0\leq f\leq 1$. From 
\cite[Thm.8.2]{Nielsen:book}, $f=1$ if and only if there are constants $\mu
_{re}$ such that 
\begin{equation}
R_{r}E_{e}C=\mu _{re}L,\quad \sum_{r,e}|\mu _{re}|^{2}=1.  \label{ind}
\end{equation}%
As shown in \cite{TaghaviKL:qec07}, maximizing the fidelity $f$ is
equivalent to minimizing the following distance, which is the approach we
focus on henceforth: 
\begin{equation}
\delta =\sum_{r,e}\left\Vert R_{r}E_{e}C-\mu _{re}L\right\Vert ^{2},
\label{dis}
\end{equation}%
where $\left\Vert X\right\Vert\equiv (\mathrm{Tr}X^{\dagger }X)^{1/2}$ is the
Frobenius norm. The minimization must now include the parameters $\{\mu
_{re}\}$ as well. Using straightforward matrix manipulations, we can rewrite
this distance as 
\begin{equation}
\delta (R,C,\Delta )=\left\Vert RE(I_{m_{E}}\otimes C)-\Delta \otimes
L\right\Vert ^{2}  \label{dis2}
\end{equation}%
where $\Delta \equiv \lbrack \mu _{re}]$ is an $m_{R}\times m_{E}$
rectangular matrix, $E$ is the $d\times m_{E}d$ rectangular error matrix
built using the error Kraus operators as horizontally arranged blocks 
\begin{equation}
E=[E_{1}...E_{m_{E}}],
\end{equation}%
and $R$ is the $m_{R}d\times d$ rectangular matrix obtained by vertically
stacking the recovery Kraus operators $R_{r}$: 
\begin{equation}
R=\left( 
\begin{array}{c}
R_{1} \\ 
\vdots  \\ 
R_{m_{R}}%
\end{array}%
\right) .
\end{equation}%
Therefore, we have $\left\Vert \Delta \right\Vert ^{2}=\mathrm{Tr}\ \Delta
^{\dagger }\Delta =\sum_{r,e}\left\vert \mu _{re}\right\vert ^{2}=1$, and $%
R^{\dagger }R=\sum_{r}R_{r}^{\dagger }R_{r}=I$.

Our optimization problem can now be summarized as follows:%
\begin{eqnarray}
&&\mathrm{minimize}\,\,\delta (R,C,\Delta )\,\,\mathrm{subject\,\,to\,\,the%
\,\,constraints}  \notag \\
R^{\dag }R &=&I,\;\;C=C^{\prime }\otimes I_{\mathrm{rec}},\ C^{\prime \dag
}C^{\prime }=I_{\mathrm{dat},\mathrm{rec}},~~\left\Vert \Delta \right\Vert
^{2}=1,  \notag \\
\label{DisOpt}
\end{eqnarray}%
where $I_{\mathrm{dat},\mathrm{rec}}$ denotes the identity operator on $%
\mathcal{H}_{\mathrm{dat}}\otimes \mathcal{H}_{\mathrm{rec}}$. There are
three matrices of parameters $(R,C,\Delta )$ in this expression that are to
be optimized. As stated, this optimization is not convex. Therefore, as
first suggested by Reimpell and Werner \cite{ReimpellW:05}, to identify the
optimal values of these parameters, we solve this optimization problem
iteratively. Starting with an arbitrary initial encoding operation, we first
find the optimized recovery operation, which by itself is a convex problem;
we then fix the recovery operation obtained in the previous step, and find
the optimized encoding operation, which is also a convex problem; and so on,
alternately optimizing the recovery and encoding operations until the
procedure converges. $\Delta $ is an intermediary matrix of parameters that
must also be recalculated at each step. This iteration continues until the
distance (\ref{dis}) stops decreasing. The details of the procedure at each
step are provided in the next two sections.

The design of the optimized recovery operation at each step is not affected
by the extra constraints added by introducing entanglement in the error
correction procedure. This is so because optimization of the recovery operation
does not depend on the details of the encoding operation. However, the
extra constraints play an important role in identifying the optimized
encoding operation, which must be of the desired form $C=C^{\prime }\otimes
I_{\mathrm{rec}}$.

The encoding ancillas must be initialized in a maximally entangled
state---for example, an EPR pair. However, the distance measure defined in (%
\ref{dis}) is independent of the initial state of the ancilla qubits. To
overcome this problem we add an extra step, an \textit{entangling} operation 
$U:\mathcal{H}_{S}\mapsto \mathcal{H}_{S}$, to the procedure. The entangling
operation does not act on the data qubits, so it can be written as $U=I_{%
\mathrm{dat}}\otimes U_{\mathrm{anc}}$, where $U_{\mathrm{anc}}:\mathcal{H}_{%
\mathrm{anc}}\mapsto \mathcal{H}_{\mathrm{anc}}$. By including this
entangling operation in the evolution, the ancillas can be assumed to have
the initial state $\left\vert 0\right\rangle $.

\section{Optimized Encoding Operation}
\label{sec:encoding}

The derivation of the optimized encoding operator is similar to what was
done previously in \cite{TaghaviKL:qec07}, but requires including the
entangling operator $U$, and a careful consideration of the subspaces
involved. Including the entangling operator, the distance measure (\ref{dis}%
) can be rewritten as%
\begin{eqnarray}
\delta (R,C,\Delta ) &=&\left\Vert RE(I_{m_{E}}\otimes CU)-\Delta \otimes
L\right\Vert ^{2}  \notag \\
&=&\sum_{r,e}\left\Vert R_{r}E_{e}CU-\mu _{re}L\right\Vert ^{2}
\notag \\
&=&\mathrm{Tr}[I+C^{\dag }C-UL^{\dag }\sum_{r,e}\mu _{re}^{\ast
  }R_{r}E_{e}C-h.c.],\notag \\
\label{del}
\end{eqnarray}%
where in the second line we used the relation between Eqs.~(\ref{dis}) and (%
\ref{dis2}), and in the third line we used the definition of the Frobenius
norm and the normalization conditions (\ref{norm}) and (\ref{ind}). 

As a first step we'll need to find the unconstrained minimum of $\delta
(R,C,\Delta )$ with respect to $C$, i.e., we need to find the solution to $%
\partial \delta /\partial C=0$ without introducing the condition $C^{\dag
}C=I$. To this end we note the following matrix differentiation identities,
valid for any pair of matrices $A$ and $Z$ with compatible dimensions, with $%
A$ being independent of the elements of $Z$: 
\begin{eqnarray}
\frac{\partial }{\partial Z}\mathrm{Tr}[AZ] &=&A^{t},\quad \frac{\partial }{%
\partial Z}\mathrm{Tr}[AZ^{\dag }]=0.  \label{d2} \\
\frac{\partial }{\partial Z}\mathrm{Tr}[ZZ^{\dag }] &=&Z^{\ast },\quad \frac{%
\partial }{\partial Z}\mathrm{Tr}[AZ^{\dag }Z]=Z^{\ast }A^{t}  \label{d4} \\
\frac{\partial }{\partial Z}\mathrm{Tr}[A(Z\otimes I)] &=&(\mathrm{Tr}%
_{2}A)^{t},  \label{d5}
\end{eqnarray}%
where in the last identity the partial trace is over the subspace acted on
by the identity operator. We prove all these identities in Appendix \ref%
{d5-proof}. 

Therefore, assuming that the recovery operators $R_{r}$ are fixed, we have 
\begin{equation}
\frac{\partial \delta }{\partial C}=C^{\ast }-(UL^{\dag }\sum_{r,e}\mu
_{re}^{\ast }R_{r}E_{e})^{t},  \label{dd/dC}
\end{equation}%
so that the solution to the unconstrained optimization problem $\frac{%
\partial \delta }{\partial C}=0$ is 
\begin{equation}
\bar{C}=\sum_{r,e}\mu _{re}(R_{r}E_{e})^{\dag }LU^{\dag }.  \label{Cbar}
\end{equation}%
We wish to impose the additional structure $C=C^{\prime }\otimes I_{\mathrm{%
rec}}$, i.e., $C^{\prime }=\frac{1}{d_{\mathrm{rec}}}\mathrm{Tr}_{\mathrm{rec%
}}C$, where the partial trace is over the recovery ancillas. Thus the
unconstrained solution over $\mathcal{H}_{\mathrm{dat}}\otimes \mathcal{H}_{%
\mathrm{enc}}$ becomes%
\begin{eqnarray}
\bar{C}^{\prime } &=&\arg \min_{C^{\prime }}\delta (R,C^{\prime }\otimes I_{%
\mathrm{rec}},\Delta )  \notag \\
&=&\frac{1}{d_{\mathrm{rec}}}\mathrm{Tr}_{\mathrm{rec}}\left[ \sum_{r,e}\mu
_{re}(R_{r}E_{e})^{\dag }L(I_{\mathrm{dat}}\otimes U_{a}^{\dag })\right] .
\label{Cbar'}
\end{eqnarray}%
The reader can check that this result can also be derived by introducing the
decomposition $C=C^{\prime }\otimes I_{\mathrm{rec}}$ directly into Eq.~(\ref%
{del}) and using identity (\ref{d5}), i.e., 
\begin{equation}
\frac{\partial \delta }{\partial C^{\prime }}=d_{\mathrm{rec}}C^{\prime \ast
}-(\mathrm{Tr}_{\mathrm{rec}}[UL^{\dag }\sum_{r,e}\mu _{re}^{\ast
}R_{r}E_{e}])^{t}.  \label{dd/dC'}
\end{equation}
This is important for the next step.

The constrained problem (\ref{DisOpt}) requires that $C^{\prime \dag
}C^{\prime }=I_{\mathrm{dat},\mathrm{rec}}$. To solve this problem we form
the Lagrangian 
\begin{equation}
\mathcal{L}=\delta (R,C^{\prime },\Delta )+\mathrm{Tr}[P(C^{\prime \dag
}C^{\prime }-I_{\mathrm{dat},\mathrm{rec}})],
\end{equation}%
where $P$ is a Hermitian Lagrange multiplier matrix (because the constraint $%
C^{\prime \dag }C^{\prime }=I_{\mathrm{dat},\mathrm{rec}}$ is Hermitian).
Setting $\partial _{C^{\prime }}\mathcal{L}=0$ we find, using Eqs.~(\ref{d4}%
), (\ref{Cbar'}), and (\ref{dd/dC'}): $C^{\prime \ast }-\bar{C}^{\prime \ast
}+C^{\prime \ast }P^{t}=0$, i.e., the solution to the constrained problem is%
\begin{equation}
C^{\prime }=\bar{C}^{\prime }(I_{\mathrm{dat},\mathrm{rec}}+P)^{-1}.
\end{equation}%
To eliminate $P$ we note that the constraint $C^{\prime \dag }C^{\prime }=I_{%
\mathrm{dat},\mathrm{rec}}$ now implies $\bar{C}^{\prime \dag }\bar{C}%
^{\prime }=(I_{\mathrm{dat},\mathrm{rec}}+P)^{2}$, from which we have $(I_{%
\mathrm{dat},\mathrm{rec}}+P)^{-1}=(\bar{C}^{\prime \dag }\bar{C}^{\prime
})^{-1/2}$. Thus, finally%
\begin{equation}
C^{\prime }=\bar{C}^{\prime }(\bar{C}^{\prime \dag }\bar{C}^{\prime
})^{-1/2},
\label{enc}
\end{equation}%
where $\bar{C}^{\prime }$ is given in Eq.~(\ref{Cbar'}). This encoding
operation is optimized for the given recovery operation $R$.

\section{Optimized Recovery Operation}
\label{recovery}

Unlike the case of the optimized
encoding operation, we do not impose any specific tensor product
structure on the recovery, so that derivation of the optimized
recovery operator is essentially the same as 
what was done previously in \cite{TaghaviKL:qec07}, except that again we
must include the entangling operator.
Here we briefly outline the steps
required to optimize the recovery operator. Using the constraints in (\ref%
{DisOpt}) and fixing the encoding operation $C$, we express the distance (%
\ref{dis}) as 
\begin{eqnarray}
\delta  &=&\left\Vert RE(I_{m_{E}}\otimes CU)-\Delta \otimes L\right\Vert
^{2}  \notag \\
&=&2d
-2\text{Re}\mathrm{Tr}RE(\Delta ^{\dagger }\otimes CUL^{\dagger }).
\label{d_recovery}
\end{eqnarray}

Minimizing (\ref{d_recovery}) with respect to $R$ is the same as maximizing
the last term. As shown in \cite[Appendix A]{TaghaviKL:qec07} this
maximization yields 
\begin{eqnarray}
&\underset{R^{\dagger }R=I}{\max }&\text{Re}\,\mathrm{Tr}\ RE(\Delta
^{\dagger }\otimes CUL^{\dagger })  \notag \\
&=&
\mathrm{Tr}\sqrt{E(\Gamma \otimes I)E^{\dagger }}
\end{eqnarray}%
where the matrix $\Gamma $ is defined as 
\begin{equation}
\Gamma =\Delta ^{\dagger }\Delta .  \label{gamma}
\end{equation}

The constraint $\left\Vert \Delta \right\Vert ^{2}=1$ from (\ref{DisOpt}) is
equivalent to $\mathrm{Tr}\ \Gamma =1$, and $\Gamma \geq 0$ by definition.
Therefore, the optimization problem for $\Gamma $ is:%
\begin{eqnarray}
&&
  \mathrm{maximize}\,\,\mathrm{Tr}\sqrt{E(\Gamma \otimes I)E^{\dag }}\,\,\mathrm{subject\,\,to}  \notag
\\
\Gamma  &\geq &0,\;\;\mathrm{Tr}\Gamma =1.  \label{OptGam}
\end{eqnarray}

The optimized $\Gamma $ can be obtained by solving an equivalent
semidefinite programming (SDP) problem, or by performing a constrained least
squares optimization \cite{TaghaviKL:qec07}. The matrix $\Delta $ can be
obtained from $\Gamma $ by the definition (\ref{gamma}) (see Eq.~(21)
of Ref.~\cite{TaghaviKL:qec07} for details). Once the optimized $%
\Delta $ is known, the optimized recovery matrix can be found from 
\begin{equation}
R=[v_{1}\ ...\ v_{d}][u_{1}\ ...\ u_{d}]^{\dagger },  \label{Rec}
\end{equation}%
where $\{v_{i},u_{i}\}_{i=1}^d$ are, respectively, the right and left singular vectors
in the singular value decomposition of the matrix $E(\Delta ^{\dagger
}\otimes CUL^{\dagger })$, with the singular values in descending order.

The algorithm below summarizes the preceding method for encoding and
recovery optimization in the presence of entanglement between the encoding
and recovery blocks.

\bigskip Initialize $C$

\textbf{Repeat}

a) Optimal recovery

\ \ \ \ \ \ maximize (\ref{OptGam}) to find $\Gamma$

\ \ \ \ \ \ solve for $\Delta $ via (\ref{gamma})

\ \ \ \ \ \ solve for $R$ via (\ref{Rec})

b) Optimal encoding

\ \ \ \ \ \ solve for $C$ via (\ref{enc})

\textbf{Until} distance (\ref{dis}) stops decreasing \bigskip

Since in each iteration of this algorithm the distance $\delta $ only
decreases, the converged solution to this optimization is guaranteed to be
at least a locally optimal solution to (\ref{DisOpt}). We apply this
procedure to a selection of different channels below to assess its
performance in practice.

\section{Random Unitary Channels}
\label{RUC}

An error channel $\mathcal{E}$ is called a \emph{random unitary channel} if
we can decompose it into the probabilistic application of one of a finite
set of unitary operations: 
\begin{equation}
\mathcal{E}(\rho )=\underset{i=1}{\overset{n}{\sum }}p_{i}V_{i}\rho
V_{i}^{\dagger },  \label{RU}
\end{equation}%
where the $V_{i}$ are unitary operators and $p_{i},i=1,\ldots ,n$ is a
probability distribution. Random unitary channels describe noise processes
that can be corrected using classical information extracted from the
environment \cite{Gregoratti:03}. It is clear that any random unitary
channel should be unital, meaning that $\mathcal{E(}I)=I$. However, the
inverse relation holds only for channels on qubits, and is not true for
higher dimensions \cite{Tregub:86,Kummerer:87}.

There are two operations in (\ref{channelU}) that are not random unitary.
These operations are the introduction of the ancillas and the partial trace
over the environment. Therefore, an error channel need not be random
unitary, of form (\ref{RU}), in general. The necessary and sufficient
conditions for a channel to be a random unitary are discussed in \cite%
{Audenaert:08}. In \cite{Buscemi:06}, an upper bound was found on the number
of unitaries needed in (\ref{RU}). Here we show that if the number of
unitary operators required in (\ref{RU}) is at most two, our optimized error
correction method can perfectly correct the error using only one EPR pair as
ancilla qubits.

In fact, for this case, the optimized encoding and recovery operations are
known, and are equivalent to teleportation. To see this, consider a channel
that can be decomposed into two unitaries, $V_{1}$ and $V_{2},$ as follows: 
\begin{equation}
\mathcal{E}(\rho )=(1-p)V_{1}\rho V_{1}^{\dagger }+pV_{2}\rho V_{2}^{\dagger
}.  \label{twoU}
\end{equation}%
Suppose Alice wants to send Bob one qubit of data through this noisy
channel. One pair of entangled qubits enables Alice and Bob to make this
communication error-free. Assume this pair of qubits was prepared by an
entanglement source and sent to Alice and Bob, each taking one of the
qubits. Alice can apply an encoding operation on her qubits---the data qubit
and her half of the ebit---before sending them to Bob through the noisy
channel. Bob applies the recovery operation to all three qubits once he
receives the message.

\begin{figure}[bp]
\includegraphics[width=3.5in,height=1.26in]{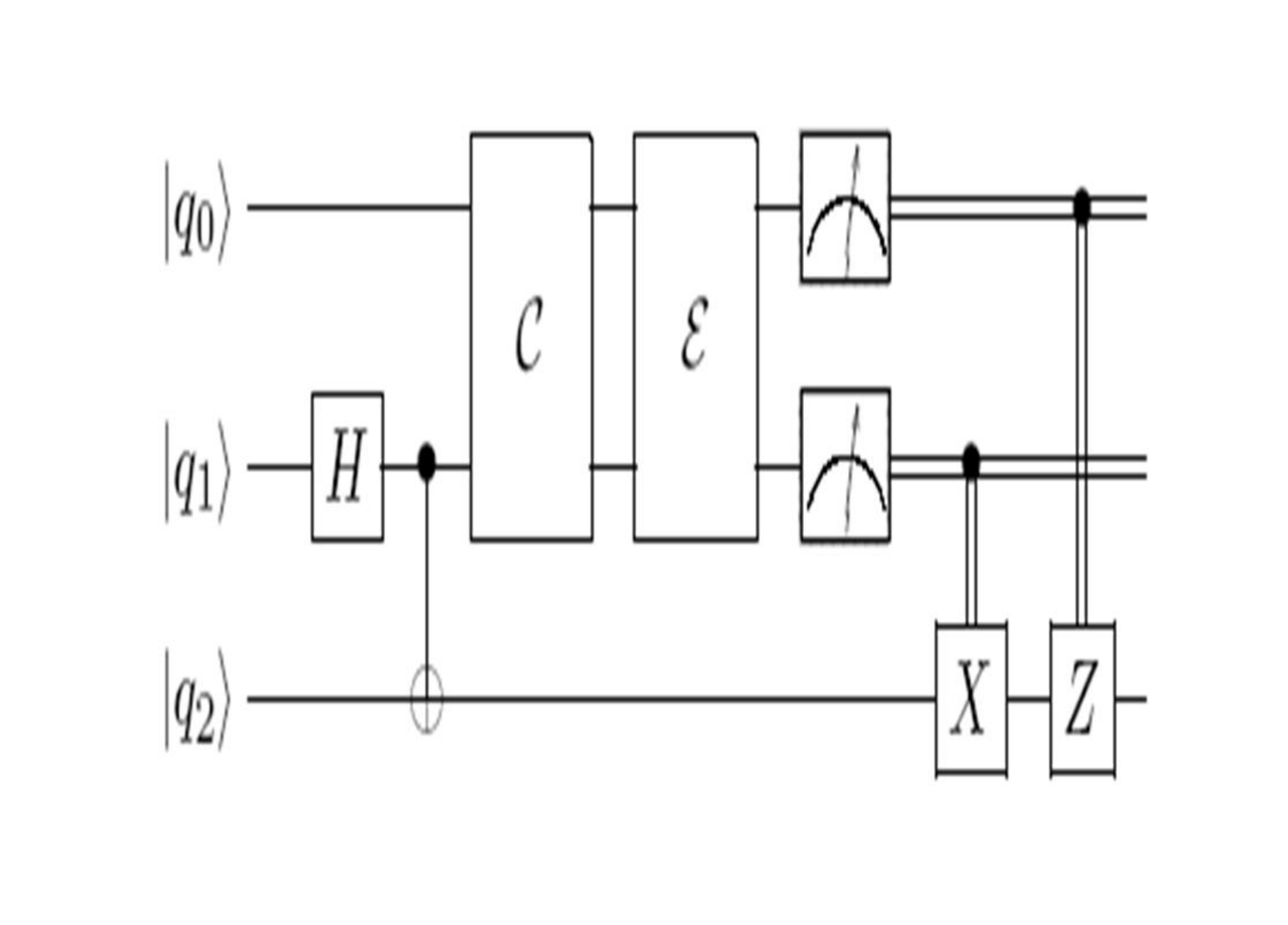}
\caption{Quantum error correction using teleportation. The information qubit 
$|q_0\rangle$ is recovered with unity fidelity in spite of passing through
the noise channel $\mathcal{E}$. The first two operations (Hadamard
gate on $|q_1\rangle$ and controlled-NOT between $|q_1\rangle$ and
$|q_2\rangle$) represent the entangling operation. This is followed by
the encoding operation $\mathcal{C}$ described by Eq.~(\protect\ref{encoding}). The last two
operations represent conditional $X$ and $Z$ gates applied to
$|q_2\rangle$, dependent on the measurement outcomes of $|q_0\rangle$
and $|q_1\rangle$. See the text for additional details.}
\label{telep}
\end{figure}

The optimized encoding operation that Alice uses in this scenario is done by
first applying $V_{1}^{\dagger }$ to her qubits (recall that we assume that
the channel is known), to make the error channel equivalent to 
\begin{equation}
\mathcal{E}(\rho )=(1-p)I\rho I+pV_{2}V_{1}^{\dagger }\rho
V_{1}V_{2}^{\dagger }  \label{oneU}
\end{equation}%
Alice continues the encoding operation by applying 
\begin{eqnarray}
C &=&\frac{1}{2}\bigl(\left\vert v_{1}v_{1}\right\rangle \left\langle
B_{1}\right\vert +\left\vert v_{1}v_{2}\right\rangle \left\langle
B_{2}\right\vert   \notag \\
&&+\left\vert v_{2}v_{1}\right\rangle \left\langle B_{3}\right\vert
+\left\vert v_{2}v_{2}\right\rangle \left\langle B_{4}\right\vert \bigr),
\label{encoding}
\end{eqnarray}%
where $v_{1}$ and $v_{2}$ are the eigenvectors of the unitary $%
V_{2}V_{1}^{\dagger }$, and $B_{1}$ to $B_{4}$ are an orthonormal basis of
maximally entangled states---for example, the Bell states $\left( \left\vert
00\right\rangle \pm \left\vert 11\right\rangle \right) /\sqrt{2}$ and $\left(
\left\vert 10\right\rangle \pm \left\vert 01\right\rangle \right) /\sqrt{2}$%
. This encoding enables us to send two bits of classical information safely
through the noisy channel. These two bits of classical information, together
with the second half of the ebit, enable us to recover the original state.

To see this in more detail suppose that the initial state of the data qubit
is $\left\vert \phi _{i}\right\rangle =a\left\vert 0\right\rangle
+b\left\vert 1\right\rangle $. The state of the entire system after applying
the entangling unitary is 
\begin{equation}
\left\vert \psi _{i}\right\rangle =\frac{1}{\sqrt{2}}[a(\left\vert
000\right\rangle +\left\vert 011\right\rangle )+b(\left\vert
100\right\rangle +\left\vert 111\right\rangle )].
\end{equation}%
After applying the encoding (\ref{encoding}) and error map on the first two
qubits, the final state of the entire system is 
\begin{eqnarray}
\left\vert \psi _{tot}\right\rangle  &=&\frac{1}{2\sqrt{2}}\bigl(\left\vert
v_{1}v_{1}\right\rangle (a\left\vert 0\right\rangle +b\left\vert
1\right\rangle )  \notag \\
&&+\left\vert v_{1}v_{2}\right\rangle (a\left\vert 0\right\rangle
-b\left\vert 1\right\rangle )  \notag \\
&&+\left\vert v_{2}v_{1}\right\rangle (a\left\vert 1\right\rangle
+b\left\vert 0\right\rangle )  \notag \\
&&+\left\vert v_{2}v_{2}\right\rangle (a\left\vert 1\right\rangle
-b\left\vert 0\right\rangle )\bigr).
\end{eqnarray}%
This is the state of the system once Bob receives it.

At this point, as shown in Figure \ref{telep}, he can recover the original
state $\left\vert \phi_{i}\right\rangle$ by measuring the state of the first
two qubits. Based on the result of the measurement he applies the
appropriate Pauli operator to the last qubit, his half of the ebit, to
recover the original state. For example, if he measures the first two qubits
to be $\left\vert v_{2}v_{1}\right\rangle$, he should apply $\sigma_x$ to
the last qubit to recover the original state.

\section{Examples}
\label{examples}

\subsection{The Bit Flip Channel}

\begin{figure}[bp]
\includegraphics[width=9cm]{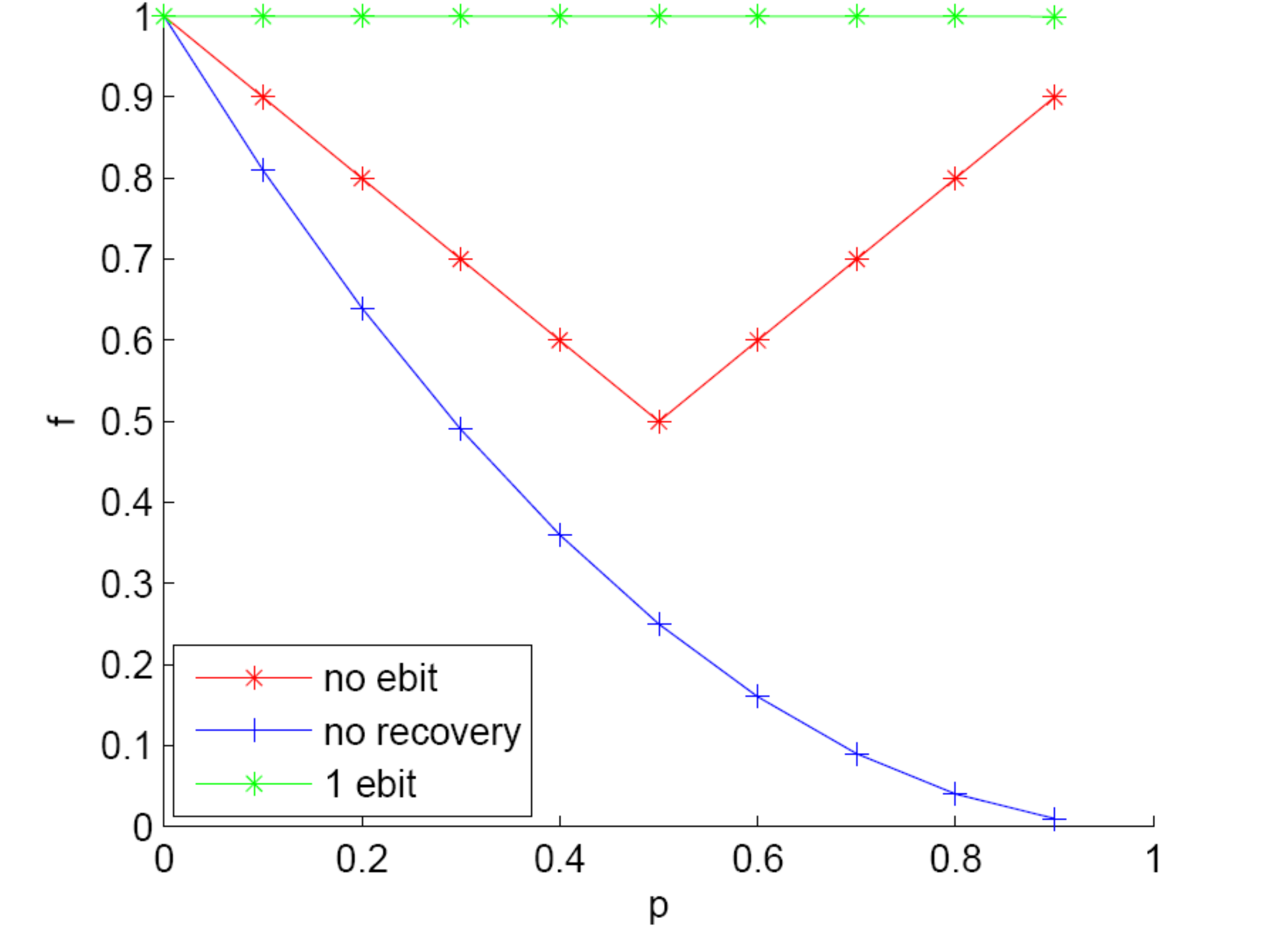}
\caption{(Color online) Optimized fidelity for the bit flip channel. Results
are shown for the case of one data qubit. The blue (bottom) curve is the
fidelity without any recovery. The red (middle) curve is the optimized
fidelity obtained for encoding using a single encoding qubit but for
recovery without entanglement, i.e., with a single recovery qubit in a
product state with the encoding qubit. This represents the standard scenario
of (channel-optimized) QEC. The green (top) curve is the fidelity obtained
in the presence of a single ebit, i.e., the EAQEC scheme shown in Fig.~%
\protect\ref{ea}. The fidelity in this case is unity for all values of the
bit flip probability $p$, showing that our optimization recovers the perfect
encoding and recovery protocol of teleportation discussed in the text.}
\label{bf}
\end{figure}

As an example of a random unitary channel with two unitary operators, we
consider the bit flip error channel, or binary symmetric channel, in which
bit flip errors occur independently with probability $p$ on each qubit. This
channel can be represented as 
\begin{equation}
\mathcal{E}(\rho )=(1-p)I\rho I+p\sigma _{x}\rho \sigma _{x},
\end{equation}%
where henceforth $\sigma _{x},\sigma _{y},\sigma _{z}$ denote the standard $%
2\times 2$ Pauli matrices. As shown in Figure \ref{bf}, this error can be
corrected using an ebit as the error-correcting resource. The ebit again is
shared between the encoding and recovery parts. The encoding acts on the
initial data qubit and the first entangled qubit (the encoding qubit).
Considering that the error is already in the form (\ref{oneU}), we do not
need to apply the initial unitary operator here. Therefore, the optimized
encoding operator based on (\ref{encoding}) is 
\begin{equation}
C=\left\vert ++\right\rangle \left\langle B_{1}\right\vert +\left\vert
+-\right\rangle \left\langle B_{2}\right\vert +\left\vert -+\right\rangle
\left\langle B_{3}\right\vert +\left\vert --\right\rangle \left\langle
B_{4}\right\vert ,
\end{equation}%
where $\left\vert +\right\rangle =(\left\vert 0\right\rangle +\left\vert
1\right\rangle )/\sqrt{2}$, $\left\vert -\right\rangle =(\left\vert
0\right\rangle -\left\vert 1\right\rangle )/\sqrt{2}$, and $B_{1}$ to $B_{4}$
form an orthonormal basis of maximally entangled states. The first entangled
qubit together with the initial data qubit are corrupted by the bit flip
error while the other entangled qubit (the recovery qubit) stays intact. The
recovery operation acts on all the qubits, and can reproduce the initial
state by measuring the first two qubits as discussed above, followed by a 
\textsc{SWAP} between the recovery qubit and the data qubit, which perfectly
restores the state of the data qubit.

While the optimization procedure we employ is not guaranteed to find the
global maximum of the channel fidelity, we see that in this case it does
indeed find the optimal solution. Convex optimization has rediscovered the
protocol of quantum teleportation.

\begin{figure}[bp]
\includegraphics[width=9cm]{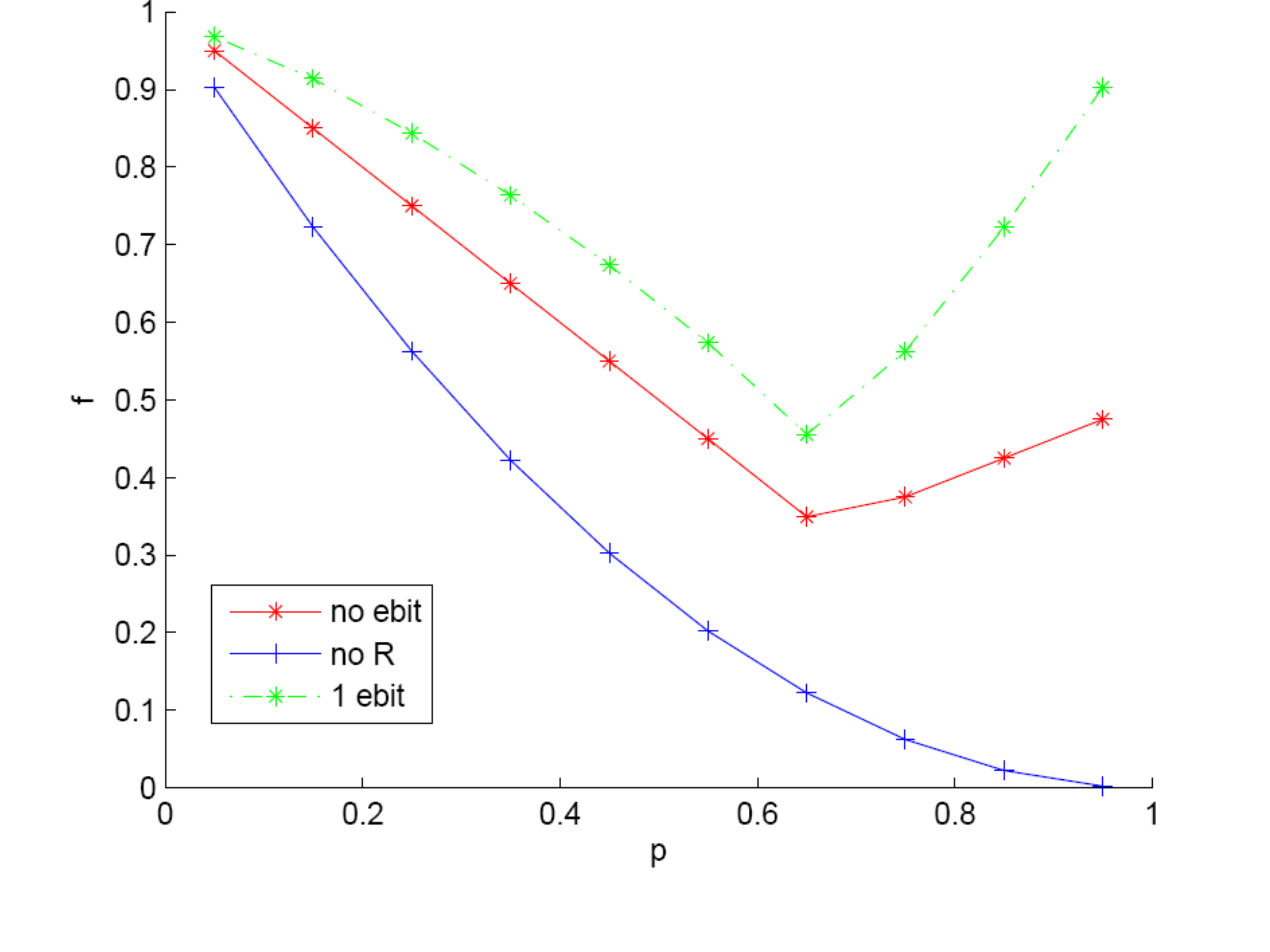} 
\caption{(Color online) Optimized fidelity for the bit-phase flip channel.
As in Fig.~\protect\ref{bf} there is a single data qubit going through the
noise channel unprotected (blue, bottom curve), optimally protected by a
single encoding and a single recovery qubit in a product state (red, middle
curve), and optimally protected by an ebit (green, top curve).}
\label{adp}
\end{figure}

\subsection{The Bit Flip/Phase Flip Channel}

In this example both bit flip and phase flip errors occur with equal
probabilities: 
\begin{equation}
\mathcal{E}(\rho ) = (1-p)I\rho I + \frac{p}{2}\sigma _{x}\rho \sigma _{x} + 
\frac{p}{2}\sigma _{z}\rho \sigma _{z}.
\end{equation}
The error occurs independently on different qubits. This error channel
clearly cannot be represented in the form of (\ref{twoU}). Hence our
teleportation argument does not apply here. There is no choice of basis for
which classical information can be sent through this channel error-free.

While entanglement cannot enable perfect error correction in this case, it
can still increase the useful information about the errors available to the
recovery block, and therefore increase the fidelity of the error correction.
Figure \ref{adp} presents the fidelity of error correction for this channel
for different values of $p$. The only difference between the upper and
middle curve is the presence of entanglement. Without entanglement, the
extra qubit used in the recovery acts as a regular recovery ancilla, and
therefore does not increase the fidelity. The minimum fidelity in both cases
occurs at $p=2/3$, where the channel is symmetric in $I$, $\sigma_x$ and $%
\sigma_z$: $\mathcal{E}(\rho ) = (1/3)(I\rho I + \sigma _{x}\rho \sigma_{x}
+ \sigma _{z}\rho \sigma _{z})$.

\begin{figure}[bp]
\includegraphics[width=9cm]{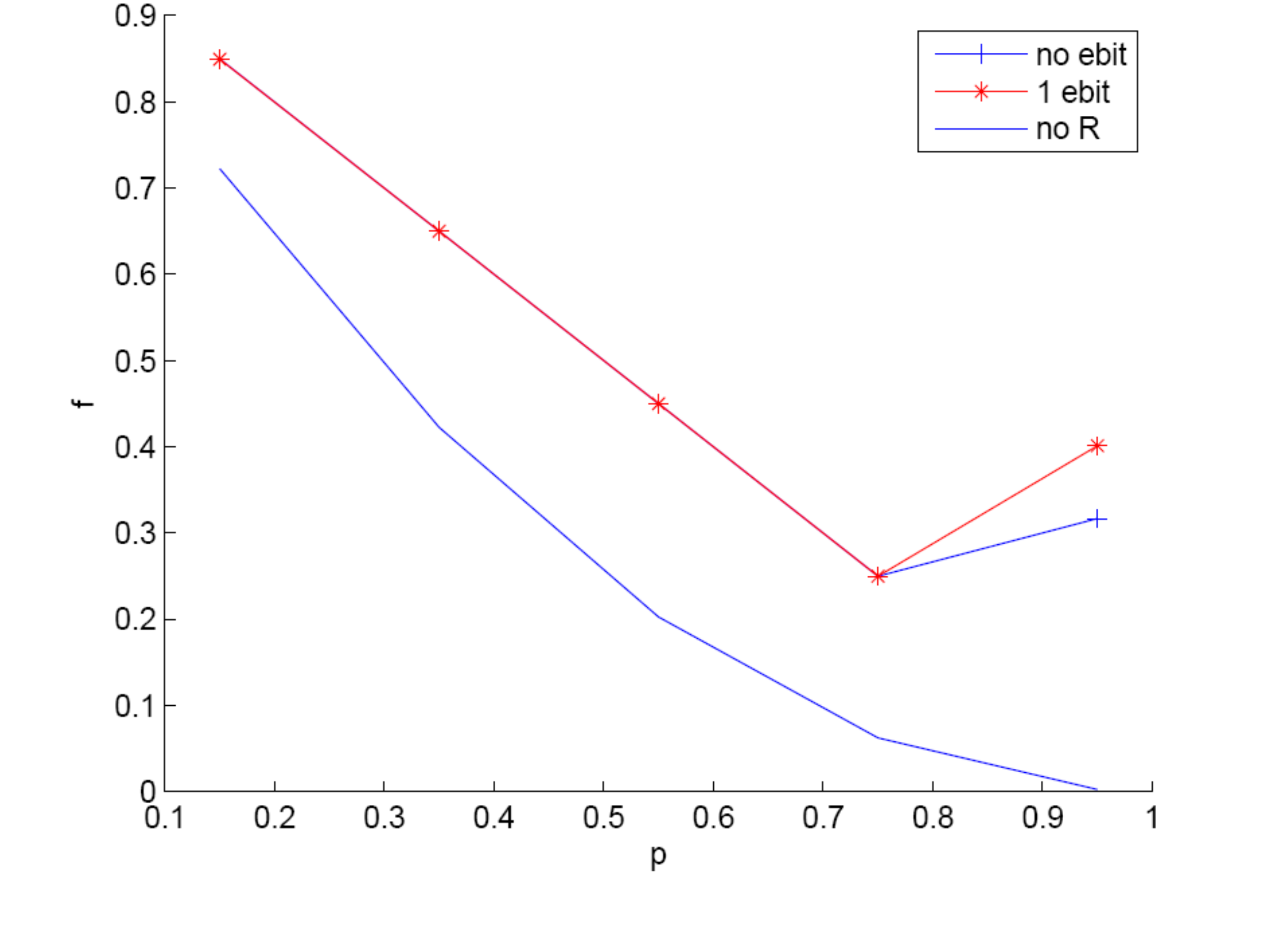}
\caption{(Color online) Optimized fidelity for the depolarizing channel. As
in Figs.~\protect\ref{bf} and \protect\ref{adp}, there is a single data
qubit going through the noise channel unprotected (blue, bottom curve),
optimally protected by a single encoding and a single recovery qubit in a
product state (blue, middle curve, overlapping red for $p\leq 3/4$), and
optimally protected by an ebit (red, top curve).}
\label{dp}
\end{figure}

\subsection{The Depolarizing Channel}

In this example we consider a particularly simple error model that is widely
used to describe noisy quantum systems: the depolarizing channel. This
important channel is often considered to be the quantum 
equivalent of the binary symmetric channel in classical error correction. It
occurs when the state of the system can be completely mixed by the action of
the channel. The noise is unbiased in the sense that it
generates bit flip errors, phase flip errors, or both with equal
probability; this is represented by the map 
\begin{equation}
\mathcal{E}(\rho )=(1-p)\rho +\frac{p}{3}\left( \sigma _{x}\rho \sigma
_{x}+\sigma _{y}\rho \sigma _{y}+\sigma _{z}\rho \sigma _{z}\right) .
\label{depolar}
\end{equation}%
The depolarizing channel shrinks the radius of the Bloch sphere by a factor
of $1-p,$ while preserving its shape \cite{Nielsen:book}.

Interestingly, for this error channel the ebit does \emph{not} increase the
fidelity for $p<3/4$. The optimized fidelity of the channel with and without
using an ebit is presented in Figure~\ref{dp}. A break point occurs at $p=3/4
$, when the coefficients in (\ref{depolar}) become equal. The fidelities of
both cases are the same for $p<3/4$, but the ebit can slightly increase the
fidelity for $p>3/4$.

An observation that may explain this difference is that for $p<3/4$ we can
write $\mathcal{E}(\rho )=(1-4p/3)\rho +(4p/3)\mathcal{T}(\rho )$, where $%
\mathcal{T}$ is the \emph{twirling} operation: 
\begin{equation}
\mathcal{T}(\rho )=\frac{1}{4}\left( \rho +\sigma _{x}\rho \sigma
_{x}+\sigma _{y}\rho \sigma _{y}+\sigma _{z}\rho \sigma _{z}\right) =\frac{1%
}{2}I,
\end{equation}%
for any valid density matrix $\rho $. In other words, with probability $4p/3$
the channel replaces the qubit with a maximally mixed qubit. For $p>3/4$
this interpretation is no longer possible, and it is in this regime that
shared entanglement seems to give improvement.

\section{Conclusion}
\label{conc}

Previous work on optimized quantum error correction
\cite{ReimpellW:05,YamamotoHT:05,FletcherSW:06,Fletcher:08,PhysRevA.77.012320,KosutLidar:06,KosutSL:07,TaghaviKL:qec07,PhysRevA.80.012326,PhysRevLett.104.120501,PhysRevA.81.062342,Tyson:09}.
considered the ``standard'', entanglement-unassisted setting. In this
work we addressed, for the first time, the problem of optimized
entanglement-assisted quantum error correction. Our methodology is a
generalization of the optimization approach developed in
Ref.~\cite{TaghaviKL:qec07}, which involves a bi-convex optimization
problem, iterating between encoding and recovery. Our main technical
innovation appears in the encoding step, where we introduced a method to
account for the constraint of initial entanglement between the encoding
and recovery ancillas.

Our method returns the optimized channel fidelity
for a given noise channel in the presence of shared entanglement between the
encoding and recovery block. We showed that entanglement can substantially
increase the error correction fidelity. For specific error channels this
fidelity increase can lead to perfect error correction. In these cases, the
error correction procedure can be interpreted as quantum teleportation. For
cases in which perfect error correction is not possible, this teleportation
interpretation does not apply; however, our optimization method returns the
optimized fidelity for these channels, in most cases showing improvement
with the resource of shared entanglement.

For one channel we found no improvement: the depolarizing channel. For the
small codes that we considered in this paper, optimized standard quantum
error correction performed just as well as optimized entanglement-assisted
quantum error correction on this channel. It is possible that this
equivalence between entanglement-assisted and standard quantum error
correction will not hold for larger codewords on the depolarizing channel.
Larger codewords would use either more ancillas or more ebits in the
encoding process. The possibility that larger codes might show a greater
difference is suggested by the fact that a three-qubit codeword with two
ebits exists that can correct an arbitrary single-qubit error \cite{Brun:06}%
, while the smallest standard code that corrects an arbitrary error is of
size five \cite{Laflamme:96}. However, the performance of \emph{optimized}
quantum error correction for larger codewords remains a subject for future
research. Larger codewords make for a more numerically intensive
optimization procedure.

Considering that regular recovery ancillas are redundant in the optimized
error correction procedure, the results of this paper show that information
about errors in the encoded data, transferred to the recovery block through
entanglement with the encoded state, can increase the fidelity for many
channels. We expect that results for a wider variety of error channels and
larger codewords will continue to bear this out.

\section*{Acknowledgments}

The authors thank Robert Kosut for insightful discussions and
comments. TAB acknowledges support from NSF Grants No.~CCF-0448658 and
No.~CCF-0830801. DAL acknowledges support from NSF Grants No.~CHM-924318,
PHY-803304, and No.~CCF-726439.

\appendix{}

\section{Trace differentiation identities}
\label{d5-proof}

We prove Eqs.~(\ref{d2}) and (\ref{d4}). These are one-line proofs which
follow directly from the definition of differentiation by a matrix \cite[%
Section 10.6]{Bernstein:book}: the derivative $\frac{\partial }{\partial Z}%
f(Z)$ of a differentiable scalar-valued function $f(Z)$ of a matrix argument 
$Z\equiv [z_{ij}]\in \mathbb{C}^{m\times n}$ is the $m\times n$ matrix whose $(i,j)$
entry is $\frac{\partial }{\partial z_{ij}}f(Z)$:

\begin{eqnarray}
\frac{\partial }{\partial Z}\mathrm{Tr}[AZ] &=&[ \frac{\partial }{%
\partial z_{ij}}\sum_{ij}z_{ij}a_{ji}] =[a_{ji}]=A^{t}, \\
\frac{\partial }{\partial Z}\mathrm{Tr}[AZ^{\dag }] &=&[ \frac{\partial 
}{\partial z_{ij}}\sum_{ij}a_{ij}z_{ij}^{\ast }] =0, \\
\frac{\partial }{\partial Z}\mathrm{Tr}[ZZ^{\dag }] &=&[ \frac{\partial 
}{\partial z_{ij}}\sum_{ij}z_{ij}z_{ij}^{\ast }] =[z_{ij}^{\ast
}]=Z^{\ast }, \\
\frac{\partial }{\partial Z}\mathrm{Tr}[AZ^{\dag }Z] &=&[ \frac{%
\partial }{\partial z_{ij}}\sum_{ij}z_{ij}(AZ^{\dag })_{ji}]
=[(AZ^{\dag })_{ji}] \notag \\
&=& Z^{\ast }A^{t},
\end{eqnarray}%
where for the second identity we used the fact that for a complex variable $%
z=x+iy$ ($x,y\in \mathbb{R}$) it holds that $\partial z^{\ast }/\partial
z=\partial (x-iy)/\partial x-i\partial (x-iy)/\partial y=0$.

Next we prove Eq.~(\ref{d5}). Let 
\begin{eqnarray}
A &=&\sum a_{mnpq}|m\rangle \langle n|\otimes |p\rangle \langle q|, \\
\tilde{Z} &=&Z\otimes I=\sum z_{m^{\prime }n^{\prime }}|m^{\prime }\rangle
\langle n^{\prime }|\otimes |p^{\prime }\rangle \langle p^{\prime }|.
\end{eqnarray}%
Then%
\begin{eqnarray}
\mathrm{Tr}A\tilde{Z} &=&\mathrm{Tr}\sum a_{mnpp^{\prime }}z_{nn^{\prime
}}|m\rangle \langle n^{\prime }|\otimes |p\rangle \langle p^{\prime }| 
\notag \\
&=&\sum_{mnp}a_{mnpp}z_{nm},
\end{eqnarray}%
so that%
\begin{eqnarray}
\frac{\partial }{\partial Z}\mathrm{Tr}[A(Z\otimes I)] &=&[ \frac{%
\partial }{\partial z_{mn}}\sum_{mnp}a_{mnpp}z_{nm}]   \notag \\
&=&[ \sum_{p}a_{nmpp}] =(\mathrm{Tr}_{2}A)^{t},
\end{eqnarray}%
where the partial trace is over the subspace acted on by the identity
operator in $Z\otimes I$.


\begin{thebibliography}{54}
\expandafter\ifx\csname natexlab\endcsname\relax\def\natexlab#1{#1}\fi
\expandafter\ifx\csname bibnamefont\endcsname\relax
  \def\bibnamefont#1{#1}\fi
\expandafter\ifx\csname bibfnamefont\endcsname\relax
  \def\bibfnamefont#1{#1}\fi
\expandafter\ifx\csname citenamefont\endcsname\relax
  \def\citenamefont#1{#1}\fi
\expandafter\ifx\csname url\endcsname\relax
  \def\url#1{\texttt{#1}}\fi
\expandafter\ifx\csname urlprefix\endcsname\relax\def\urlprefix{URL }\fi
\providecommand{\bibinfo}[2]{#2}
\providecommand{\eprint}[2][]{\url{#2}}

\bibitem[{\citenamefont{{P.W. Shor}}(1995)}]{Shor:95}
\bibinfo{author}{\bibnamefont{{P.W. Shor}}}, \bibinfo{journal}{Phys. Rev. A}
  \textbf{\bibinfo{volume}{52}}, \bibinfo{pages}{{R2493}}
  (\bibinfo{year}{1995}).

\bibitem[{\citenamefont{{A.M. Steane}}(1996)}]{Steane:96a}
\bibinfo{author}{\bibnamefont{{A.M. Steane}}}, \bibinfo{journal}{Phys. Rev.
  Lett.} \textbf{\bibinfo{volume}{77}}, \bibinfo{pages}{793}
  (\bibinfo{year}{1996}).

\bibitem[{\citenamefont{Gottesman}(1996)}]{Gottesman:96}
\bibinfo{author}{\bibfnamefont{D.}~\bibnamefont{Gottesman}},
  \bibinfo{journal}{Phys. Rev. A} \textbf{\bibinfo{volume}{54}},
  \bibinfo{pages}{1862} (\bibinfo{year}{1996}).

\bibitem[{\citenamefont{{E. Knill and R. Laflamme}}(1997)}]{Knill:97b}
\bibinfo{author}{\bibnamefont{{E. Knill and R. Laflamme}}},
  \bibinfo{journal}{Phys. Rev. A} \textbf{\bibinfo{volume}{55}},
  \bibinfo{pages}{900} (\bibinfo{year}{1997}).

\bibitem[{\citenamefont{{D. Aharonov and M. Ben-Or}}(1997)}]{Aharonov:96}
\bibinfo{author}{\bibnamefont{{D. Aharonov and M. Ben-Or}}}, in
  \emph{\bibinfo{booktitle}{{Proceedings of 29th Annual ACM Symposium on Theory
  of Computing (STOC)}}} (\bibinfo{publisher}{{ACM}}, \bibinfo{address}{{New
  York, NY}}, \bibinfo{year}{1997}), p. \bibinfo{pages}{176}.

\bibitem[{\citenamefont{{A.Yu. Kitaev}}(1996)}]{Kitaev:96}
\bibinfo{author}{\bibnamefont{{A.Yu. Kitaev}}}, \bibinfo{journal}{Russian Math.
  Surveys} \textbf{\bibinfo{volume}{52}}, \bibinfo{pages}{1191}
  (\bibinfo{year}{1996}).

\bibitem[{\citenamefont{Gottesman}(1998)}]{Gottesman:97a}
\bibinfo{author}{\bibfnamefont{D.}~\bibnamefont{Gottesman}},
  \bibinfo{journal}{Phys. Rev. A} \textbf{\bibinfo{volume}{57}},
  \bibinfo{pages}{127} (\bibinfo{year}{1998}).

\bibitem[{\citenamefont{Preskill}(1998)}]{Preskill:97a}
\bibinfo{author}{\bibfnamefont{J.}~\bibnamefont{Preskill}},
  \bibinfo{journal}{Proc. R. Soc. London Ser. A}
  \textbf{\bibinfo{volume}{454}}, \bibinfo{pages}{385} (\bibinfo{year}{1998}).

\bibitem[{\citenamefont{Bennett et~al.}(1996)\citenamefont{Bennett, DiVincenzo,
  Smolin, and Wootters}}]{Breeding:96}
\bibinfo{author}{\bibfnamefont{C.}~\bibnamefont{Bennett}},
  \bibinfo{author}{\bibfnamefont{D.}~\bibnamefont{DiVincenzo}},
  \bibinfo{author}{\bibfnamefont{J.}~\bibnamefont{Smolin}}, \bibnamefont{and}
  \bibinfo{author}{\bibfnamefont{W.}~\bibnamefont{Wootters}},
  \bibinfo{journal}{Phys. Rev. A} \textbf{\bibinfo{volume}{54}},
  \bibinfo{pages}{3824} (\bibinfo{year}{1996}).

\bibitem[{\citenamefont{Bowen}(2002)}]{Bowen:02}
\bibinfo{author}{\bibfnamefont{G.}~\bibnamefont{Bowen}},
  \bibinfo{journal}{Phys. Rev. A} \textbf{\bibinfo{volume}{66}},
  \bibinfo{pages}{052313} (\bibinfo{year}{2002}).

\bibitem[{\citenamefont{Brun et~al.}(2006{\natexlab{a}})\citenamefont{Brun,
  Devetak, and Hsieh}}]{Brun:06}
\bibinfo{author}{\bibfnamefont{T.}~\bibnamefont{Brun}},
  \bibinfo{author}{\bibfnamefont{I.}~\bibnamefont{Devetak}}, \bibnamefont{and}
  \bibinfo{author}{\bibfnamefont{M.-H.} \bibnamefont{Hsieh}},
  \bibinfo{journal}{Science} \textbf{\bibinfo{volume}{314}},
  \bibinfo{pages}{436} (\bibinfo{year}{2006}{\natexlab{a}}).

\bibitem[{\citenamefont{Brun et~al.}(2006{\natexlab{b}})\citenamefont{Brun,
  Devetak, and Hsieh}}]{Brun:06b}
\bibinfo{author}{\bibfnamefont{T.}~\bibnamefont{Brun}},
  \bibinfo{author}{\bibfnamefont{I.}~\bibnamefont{Devetak}}, \bibnamefont{and}
  \bibinfo{author}{\bibfnamefont{M.-H.} \bibnamefont{Hsieh}}
  (\bibinfo{year}{2006}{\natexlab{b}}), \bibinfo{note}{eprint
  quant-ph/0608027}.

\bibitem[{\citenamefont{Hsieh et~al.}(2007)\citenamefont{Hsieh, Devetak, and
  Brun}}]{Hsieh:07}
\bibinfo{author}{\bibfnamefont{M.-H.} \bibnamefont{Hsieh}},
  \bibinfo{author}{\bibfnamefont{I.}~\bibnamefont{Devetak}}, \bibnamefont{and}
  \bibinfo{author}{\bibfnamefont{T.}~\bibnamefont{Brun}},
  \bibinfo{journal}{Phys. Rev. A} \textbf{\bibinfo{volume}{76}},
  \bibinfo{pages}{062313} (\bibinfo{year}{2007}).

\bibitem[{\citenamefont{Wilde and Brun}(2008)}]{Wilde:08}
\bibinfo{author}{\bibfnamefont{M.}~\bibnamefont{Wilde}} \bibnamefont{and}
  \bibinfo{author}{\bibfnamefont{T.}~\bibnamefont{Brun}},
  \bibinfo{journal}{Phys. Rev. A} \textbf{\bibinfo{volume}{77}},
  \bibinfo{pages}{064302} (\bibinfo{year}{2008}).

\bibitem[{\citenamefont{Kremsky et~al.}(2008)\citenamefont{Kremsky, Hsieh, and
  Brun}}]{Kremsky:08}
\bibinfo{author}{\bibfnamefont{I.}~\bibnamefont{Kremsky}},
  \bibinfo{author}{\bibfnamefont{M.-H.} \bibnamefont{Hsieh}}, \bibnamefont{and}
  \bibinfo{author}{\bibfnamefont{T.}~\bibnamefont{Brun}},
  \bibinfo{journal}{Phys. Rev. A} \textbf{\bibinfo{volume}{78}},
  \bibinfo{pages}{012341} (\bibinfo{year}{2008}).

\bibitem[{\citenamefont{Hsieh et~al.}(2009)\citenamefont{Hsieh, Devetak, and
  Brun}}]{Hsieh:09}
\bibinfo{author}{\bibfnamefont{M.-H.} \bibnamefont{Hsieh}},
  \bibinfo{author}{\bibfnamefont{I.}~\bibnamefont{Devetak}}, \bibnamefont{and}
  \bibinfo{author}{\bibfnamefont{T.}~\bibnamefont{Brun}},
  \bibinfo{journal}{Phys. Rev. A} \textbf{\bibinfo{volume}{79}},
  \bibinfo{pages}{032340} (\bibinfo{year}{2009}).

\bibitem[{\citenamefont{Wilde and Brun}(2010)}]{Wilde:10}
\bibinfo{author}{\bibfnamefont{M.}~\bibnamefont{Wilde}} \bibnamefont{and}
  \bibinfo{author}{\bibfnamefont{T.}~\bibnamefont{Brun}},
  \bibinfo{journal}{Phys. Rev. A} \textbf{\bibinfo{volume}{81}},
  \bibinfo{pages}{042333} (\bibinfo{year}{2010}).

\bibitem[{\citenamefont{Altepeter et~al.}(2003)\citenamefont{Altepeter,
  Branning, Jeffrey, Wei, Kwiat, Thew, O'Brien, Nielsen, and
  White}}]{PhysRevLett.90.193601}
\bibinfo{author}{\bibfnamefont{J.~B.} \bibnamefont{Altepeter}},
  \bibinfo{author}{\bibfnamefont{D.}~\bibnamefont{Branning}},
  \bibinfo{author}{\bibfnamefont{E.}~\bibnamefont{Jeffrey}},
  \bibinfo{author}{\bibfnamefont{T.~C.} \bibnamefont{Wei}},
  \bibinfo{author}{\bibfnamefont{P.~G.} \bibnamefont{Kwiat}},
  \bibinfo{author}{\bibfnamefont{R.~T.} \bibnamefont{Thew}},
  \bibinfo{author}{\bibfnamefont{J.~L.} \bibnamefont{O'Brien}},
  \bibinfo{author}{\bibfnamefont{M.~A.} \bibnamefont{Nielsen}},
  \bibnamefont{and} \bibinfo{author}{\bibfnamefont{A.~G.} \bibnamefont{White}},
  \bibinfo{journal}{Phys. Rev. Lett.} \textbf{\bibinfo{volume}{90}},
  \bibinfo{pages}{193601} (\bibinfo{year}{2003}).

\bibitem[{\citenamefont{Weinstein et~al.}(2004)\citenamefont{Weinstein, Havel,
  Emerson, Boulant, Saraceno, Lloyd, and Cory}}]{Weinstein:04}
\bibinfo{author}{\bibfnamefont{Y.}~\bibnamefont{Weinstein}},
  \bibinfo{author}{\bibfnamefont{T.}~\bibnamefont{Havel}},
  \bibinfo{author}{\bibfnamefont{J.}~\bibnamefont{Emerson}},
  \bibinfo{author}{\bibfnamefont{N.}~\bibnamefont{Boulant}},
  \bibinfo{author}{\bibfnamefont{M.}~\bibnamefont{Saraceno}},
  \bibinfo{author}{\bibfnamefont{S.}~\bibnamefont{Lloyd}}, \bibnamefont{and}
  \bibinfo{author}{\bibfnamefont{D.}~\bibnamefont{Cory}}, \bibinfo{journal}{J.
  Chem. Phys.} \textbf{\bibinfo{volume}{121}}, \bibinfo{pages}{6117}
  (\bibinfo{year}{2004}).

\bibitem[{\citenamefont{Howard et~al.}(2006)\citenamefont{Howard, Twamley,
  Wittman, Gaebel, Jelezko, and Wrachtrup}}]{Howard:06}
\bibinfo{author}{\bibfnamefont{M.}~\bibnamefont{Howard}},
  \bibinfo{author}{\bibfnamefont{J.}~\bibnamefont{Twamley}},
  \bibinfo{author}{\bibfnamefont{C.}~\bibnamefont{Wittman}},
  \bibinfo{author}{\bibfnamefont{T.}~\bibnamefont{Gaebel}},
  \bibinfo{author}{\bibfnamefont{F.}~\bibnamefont{Jelezko}}, \bibnamefont{and}
  \bibinfo{author}{\bibfnamefont{J.}~\bibnamefont{Wrachtrup}},
  \bibinfo{journal}{New J. Phys.} \textbf{\bibinfo{volume}{8}},
  \bibinfo{pages}{33} (\bibinfo{year}{2006}).

\bibitem[{\citenamefont{Chuang and Nielsen}(1997)}]{Chuang:97c}
\bibinfo{author}{\bibfnamefont{I.}~\bibnamefont{Chuang}} \bibnamefont{and}
  \bibinfo{author}{\bibfnamefont{M.}~\bibnamefont{Nielsen}},
  \bibinfo{journal}{J. Mod. Optics} \textbf{\bibinfo{volume}{44}},
  \bibinfo{pages}{2455} (\bibinfo{year}{1997}).

\bibitem[{\citenamefont{Poyatos et~al.}(1997)\citenamefont{Poyatos, Cirac, and
  Zoller}}]{Poyatos:97}
\bibinfo{author}{\bibfnamefont{J.}~\bibnamefont{Poyatos}},
  \bibinfo{author}{\bibfnamefont{J.}~\bibnamefont{Cirac}}, \bibnamefont{and}
  \bibinfo{author}{\bibfnamefont{P.}~\bibnamefont{Zoller}},
  \bibinfo{journal}{Phys. Rev. Lett.} \textbf{\bibinfo{volume}{78}},
  \bibinfo{pages}{390} (\bibinfo{year}{1997}).

\bibitem[{\citenamefont{D'Ariano and Presti}(2001)}]{DAriano:01}
\bibinfo{author}{\bibfnamefont{G.~M.} \bibnamefont{D'Ariano}} \bibnamefont{and}
  \bibinfo{author}{\bibfnamefont{P.~L.} \bibnamefont{Presti}},
  \bibinfo{journal}{Phys. Rev. Lett.} \textbf{\bibinfo{volume}{86}},
  \bibinfo{pages}{4195} (\bibinfo{year}{2001}).

\bibitem[{\citenamefont{Mohseni and Lidar}(2006)}]{MohseniLidar:06}
\bibinfo{author}{\bibfnamefont{M.}~\bibnamefont{Mohseni}} \bibnamefont{and}
  \bibinfo{author}{\bibfnamefont{D.~A.} \bibnamefont{Lidar}},
  \bibinfo{journal}{Phys. Rev. Lett.} \textbf{\bibinfo{volume}{97}},
  \bibinfo{pages}{170501} (\bibinfo{year}{2006}).

\bibitem[{\citenamefont{Emerson et~al.}(2007)\citenamefont{Emerson, Silva,
  Moussa, Ryan, Laforest, Baugh, Cory, and Laflamme}}]{EmersonScience:07}
\bibinfo{author}{\bibfnamefont{J.}~\bibnamefont{Emerson}},
  \bibinfo{author}{\bibfnamefont{M.}~\bibnamefont{Silva}},
  \bibinfo{author}{\bibfnamefont{O.}~\bibnamefont{Moussa}},
  \bibinfo{author}{\bibfnamefont{C.}~\bibnamefont{Ryan}},
  \bibinfo{author}{\bibfnamefont{M.}~\bibnamefont{Laforest}},
  \bibinfo{author}{\bibfnamefont{J.}~\bibnamefont{Baugh}},
  \bibinfo{author}{\bibfnamefont{D.~G.} \bibnamefont{Cory}}, \bibnamefont{and}
  \bibinfo{author}{\bibfnamefont{R.}~\bibnamefont{Laflamme}},
  \bibinfo{journal}{Science} \textbf{\bibinfo{volume}{317}},
  \bibinfo{pages}{1893} (\bibinfo{year}{2007}).

\bibitem[{\citenamefont{{M. Mohseni, A. T. Rezakhani, and D. A.
  Lidar}}(2008)}]{MohseniRezakhaniLidar}
\bibinfo{author}{\bibnamefont{{M. Mohseni, A. T. Rezakhani, and D. A. Lidar}}},
  \bibinfo{journal}{Phys. Rev. A} \textbf{\bibinfo{volume}{77}},
  \bibinfo{pages}{032322} (\bibinfo{year}{2008}).

\bibitem[{\citenamefont{Branderhorst et~al.}(2009)\citenamefont{Branderhorst,
  Nunn, Walmsley, and Kosut}}]{Branderhorst:09}
\bibinfo{author}{\bibfnamefont{M.}~\bibnamefont{Branderhorst}},
  \bibinfo{author}{\bibfnamefont{J.}~\bibnamefont{Nunn}},
  \bibinfo{author}{\bibfnamefont{I.}~\bibnamefont{Walmsley}}, \bibnamefont{and}
  \bibinfo{author}{\bibfnamefont{R.}~\bibnamefont{Kosut}},
  \bibinfo{journal}{New J. Phys.} \textbf{\bibinfo{volume}{11}},
  \bibinfo{pages}{115010} (\bibinfo{year}{2009}).

\bibitem[{\citenamefont{Reimpell and Werner}(2005)}]{ReimpellW:05}
\bibinfo{author}{\bibfnamefont{M.}~\bibnamefont{Reimpell}} \bibnamefont{and}
  \bibinfo{author}{\bibfnamefont{R.~F.} \bibnamefont{Werner}},
  \bibinfo{journal}{Phys. Rev. Lett.} \textbf{\bibinfo{volume}{94}},
  \bibinfo{pages}{080501} (\bibinfo{year}{2005}).

\bibitem[{\citenamefont{Yamamoto et~al.}(2005)\citenamefont{Yamamoto, Hara, and
  Tsumura}}]{YamamotoHT:05}
\bibinfo{author}{\bibfnamefont{N.}~\bibnamefont{Yamamoto}},
  \bibinfo{author}{\bibfnamefont{S.}~\bibnamefont{Hara}}, \bibnamefont{and}
  \bibinfo{author}{\bibfnamefont{K.}~\bibnamefont{Tsumura}},
  \bibinfo{journal}{Phys. Rev. A} \textbf{\bibinfo{volume}{71}},
  \bibinfo{pages}{022322} (\bibinfo{year}{2005}).

\bibitem[{\citenamefont{Fletcher et~al.}(2007)\citenamefont{Fletcher, Shor, and
  Win}}]{FletcherSW:06}
\bibinfo{author}{\bibfnamefont{A.~S.} \bibnamefont{Fletcher}},
  \bibinfo{author}{\bibfnamefont{P.~W.} \bibnamefont{Shor}}, \bibnamefont{and}
  \bibinfo{author}{\bibfnamefont{M.~Z.} \bibnamefont{Win}},
  \bibinfo{journal}{Phys. Rev. A} \textbf{\bibinfo{volume}{75}},
  \bibinfo{pages}{012338} (\bibinfo{year}{2007}).

\bibitem[{\citenamefont{Fletcher
  et~al.}(2008{\natexlab{a}})\citenamefont{Fletcher, Shor, and
  Win}}]{Fletcher:08}
\bibinfo{author}{\bibfnamefont{A.}~\bibnamefont{Fletcher}},
  \bibinfo{author}{\bibfnamefont{P.}~\bibnamefont{Shor}}, \bibnamefont{and}
  \bibinfo{author}{\bibfnamefont{M.}~\bibnamefont{Win}}, \bibinfo{journal}{IEEE
  Trans. Inf. Theory} \textbf{\bibinfo{volume}{54}}, \bibinfo{pages}{5705}
  (\bibinfo{year}{2008}{\natexlab{a}}).

\bibitem[{\citenamefont{Fletcher
  et~al.}(2008{\natexlab{b}})\citenamefont{Fletcher, Shor, and
  Win}}]{PhysRevA.77.012320}
\bibinfo{author}{\bibfnamefont{A.~S.} \bibnamefont{Fletcher}},
  \bibinfo{author}{\bibfnamefont{P.~W.} \bibnamefont{Shor}}, \bibnamefont{and}
  \bibinfo{author}{\bibfnamefont{M.~Z.} \bibnamefont{Win}},
  \bibinfo{journal}{Phys. Rev. A} \textbf{\bibinfo{volume}{77}},
  \bibinfo{pages}{012320} (\bibinfo{year}{2008}{\natexlab{b}}).

\bibitem[{\citenamefont{Kosut and Lidar}(2009)}]{KosutLidar:06}
\bibinfo{author}{\bibfnamefont{R.}~\bibnamefont{Kosut}} \bibnamefont{and}
  \bibinfo{author}{\bibfnamefont{D.~A.} \bibnamefont{Lidar}},
  \bibinfo{journal}{Quant. Inf. Proc.} \textbf{\bibinfo{volume}{8}},
  \bibinfo{pages}{441} (\bibinfo{year}{2009}).

\bibitem[{\citenamefont{Kosut et~al.}(2008)\citenamefont{Kosut, Shabani, and
  Lidar}}]{KosutSL:07}
\bibinfo{author}{\bibfnamefont{R.}~\bibnamefont{Kosut}},
  \bibinfo{author}{\bibfnamefont{A.}~\bibnamefont{Shabani}}, \bibnamefont{and}
  \bibinfo{author}{\bibfnamefont{D.~A.} \bibnamefont{Lidar}},
  \bibinfo{journal}{Phys. Rev. Lett.} \textbf{\bibinfo{volume}{100}},
  \bibinfo{pages}{020502} (\bibinfo{year}{2008}).

\bibitem[{\citenamefont{Taghavi et~al.}(2010)\citenamefont{Taghavi, Kosut, and
  Lidar}}]{TaghaviKL:qec07}
\bibinfo{author}{\bibfnamefont{S.}~\bibnamefont{Taghavi}},
  \bibinfo{author}{\bibfnamefont{R.~L.} \bibnamefont{Kosut}}, \bibnamefont{and}
  \bibinfo{author}{\bibfnamefont{D.~A.} \bibnamefont{Lidar}},
  \bibinfo{journal}{IEEE Trans. Inf. Theory} \textbf{\bibinfo{volume}{56}},
  \bibinfo{pages}{1461} (\bibinfo{year}{2010}).

\bibitem[{\citenamefont{Ball\'o and Gurin}(2009)}]{PhysRevA.80.012326}
\bibinfo{author}{\bibfnamefont{G.}~\bibnamefont{Ball\'o}} \bibnamefont{and}
  \bibinfo{author}{\bibfnamefont{P.}~\bibnamefont{Gurin}},
  \bibinfo{journal}{Phys. Rev. A} \textbf{\bibinfo{volume}{80}},
  \bibinfo{pages}{012326} (\bibinfo{year}{2009}).

\bibitem[{\citenamefont{B\'eny and Oreshkov}(2010)}]{PhysRevLett.104.120501}
\bibinfo{author}{\bibfnamefont{C.}~\bibnamefont{B\'eny}} \bibnamefont{and}
  \bibinfo{author}{\bibfnamefont{O.}~\bibnamefont{Oreshkov}},
  \bibinfo{journal}{Phys. Rev. Lett.} \textbf{\bibinfo{volume}{104}},
  \bibinfo{pages}{120501} (\bibinfo{year}{2010}).

\bibitem[{\citenamefont{Ng and Mandayam}(2010)}]{PhysRevA.81.062342}
\bibinfo{author}{\bibfnamefont{H.~K.} \bibnamefont{Ng}} \bibnamefont{and}
  \bibinfo{author}{\bibfnamefont{P.}~\bibnamefont{Mandayam}},
  \bibinfo{journal}{Phys. Rev. A} \textbf{\bibinfo{volume}{81}},
  \bibinfo{pages}{062342} (\bibinfo{year}{2010}).

\bibitem[{\citenamefont{Tyson}(2009)}]{Tyson:09}
\bibinfo{author}{\bibfnamefont{J.}~\bibnamefont{Tyson}} (\bibinfo{year}{2009}),
  \bibinfo{note}{eprint arXiv:0907.3386}.

\bibitem[{\citenamefont{Shabani and Lidar}(2009)}]{ShabaniLidar:08}
\bibinfo{author}{\bibfnamefont{A.}~\bibnamefont{Shabani}} \bibnamefont{and}
  \bibinfo{author}{\bibfnamefont{D.~A.} \bibnamefont{Lidar}},
  \bibinfo{journal}{Phys. Rev. Lett.} \textbf{\bibinfo{volume}{102}},
  \bibinfo{pages}{100402} (\bibinfo{year}{2009}).

\bibitem[{\citenamefont{Ollivier and Zurek}(2002)}]{Ollivier:01}
\bibinfo{author}{\bibfnamefont{H.}~\bibnamefont{Ollivier}} \bibnamefont{and}
  \bibinfo{author}{\bibfnamefont{W.}~\bibnamefont{Zurek}},
  \bibinfo{journal}{Phys. Rev. Lett.} \textbf{\bibinfo{volume}{88}},
  \bibinfo{pages}{017901} (\bibinfo{year}{2002}).

\bibitem[{\citenamefont{Rodr\'{i}guez-Rosario
  et~al.}(2008)\citenamefont{Rodr\'{i}guez-Rosario, Modi, Kuah, Sudarshan, and
  Shaji}}]{Rodriguez:07}
\bibinfo{author}{\bibfnamefont{C.}~\bibnamefont{Rodr\'{i}guez-Rosario}},
  \bibinfo{author}{\bibfnamefont{K.}~\bibnamefont{Modi}},
  \bibinfo{author}{\bibfnamefont{A.-M.} \bibnamefont{Kuah}},
  \bibinfo{author}{\bibfnamefont{E.}~\bibnamefont{Sudarshan}},
  \bibnamefont{and} \bibinfo{author}{\bibfnamefont{A.}~\bibnamefont{Shaji}},
  \bibinfo{journal}{J. Phys. A} \textbf{\bibinfo{volume}{41}},
  \bibinfo{pages}{205301} (\bibinfo{year}{2008}).

\bibitem[{\citenamefont{Kraus}(1983)}]{Kraus:83}
\bibinfo{author}{\bibfnamefont{K.}~\bibnamefont{Kraus}},
  \emph{\bibinfo{title}{States, Effects and Operations}}, Fundamental Notions
  of Quantum Theory (\bibinfo{publisher}{Academic}, \bibinfo{address}{Berlin},
  \bibinfo{year}{1983}).

\bibitem[{\citenamefont{Nielsen and Chuang}(2000)}]{Nielsen:book}
\bibinfo{author}{\bibfnamefont{M.}~\bibnamefont{Nielsen}} \bibnamefont{and}
  \bibinfo{author}{\bibfnamefont{I.}~\bibnamefont{Chuang}},
  \emph{\bibinfo{title}{Quantum Computation and Quantum Information}}
  (\bibinfo{publisher}{Cambridge University Press},
  \bibinfo{address}{Cambridge, England}, \bibinfo{year}{2000}).

\bibitem[{\citenamefont{Breuer and Petruccione}(2002)}]{Breuer:book}
\bibinfo{author}{\bibfnamefont{H.-P.} \bibnamefont{Breuer}} \bibnamefont{and}
  \bibinfo{author}{\bibfnamefont{F.}~\bibnamefont{Petruccione}},
  \emph{\bibinfo{title}{The Theory of Open Quantum Systems}}
  (\bibinfo{publisher}{Oxford University Press}, \bibinfo{address}{Oxford},
  \bibinfo{year}{2002}).

\bibitem[{\citenamefont{Gilchrist et~al.}(2005)\citenamefont{Gilchrist,
  Langford, and Nielsen}}]{GilchristLN:04}
\bibinfo{author}{\bibfnamefont{A.}~\bibnamefont{Gilchrist}},
  \bibinfo{author}{\bibfnamefont{N.~K.} \bibnamefont{Langford}},
  \bibnamefont{and} \bibinfo{author}{\bibfnamefont{M.~A.}
  \bibnamefont{Nielsen}}, \bibinfo{journal}{Phys. Rev. A}
  \textbf{\bibinfo{volume}{71}}, \bibinfo{pages}{062310}
  (\bibinfo{year}{2005}).

\bibitem[{\citenamefont{Kretschmann and Werner}(2004)}]{Kretschmann:04}
\bibinfo{author}{\bibfnamefont{D.}~\bibnamefont{Kretschmann}} \bibnamefont{and}
  \bibinfo{author}{\bibfnamefont{R.~F.} \bibnamefont{Werner}},
  \bibinfo{journal}{New J. Phys.} \textbf{\bibinfo{volume}{6}},
  \bibinfo{pages}{26} (\bibinfo{year}{2004}).

\bibitem[{\citenamefont{Gregoratti and Werner}(2003)}]{Gregoratti:03}
\bibinfo{author}{\bibfnamefont{M.}~\bibnamefont{Gregoratti}} \bibnamefont{and}
  \bibinfo{author}{\bibfnamefont{R.}~\bibnamefont{Werner}},
  \bibinfo{journal}{J. Mod. Optics} \textbf{\bibinfo{volume}{50}},
  \bibinfo{pages}{915} (\bibinfo{year}{2003}).

\bibitem[{\citenamefont{Tregub}(1986)}]{Tregub:86}
\bibinfo{author}{\bibfnamefont{S.~L.} \bibnamefont{Tregub}},
  \bibinfo{journal}{Sov. Math.} \textbf{\bibinfo{volume}{30}},
  \bibinfo{pages}{105} (\bibinfo{year}{1986}).

\bibitem[{\citenamefont{K\"{u}mmerer and Maassen}(1987)}]{Kummerer:87}
\bibinfo{author}{\bibfnamefont{B.}~\bibnamefont{K\"{u}mmerer}}
  \bibnamefont{and} \bibinfo{author}{\bibfnamefont{H.}~\bibnamefont{Maassen}},
  \bibinfo{journal}{Commun. Math. Phys.} \textbf{\bibinfo{volume}{109}},
  \bibinfo{pages}{1} (\bibinfo{year}{1987}).

\bibitem[{\citenamefont{Audenaert and Scheel}(2008)}]{Audenaert:08}
\bibinfo{author}{\bibfnamefont{K.~M.~R.} \bibnamefont{Audenaert}}
  \bibnamefont{and} \bibinfo{author}{\bibfnamefont{S.}~\bibnamefont{Scheel}},
  \bibinfo{journal}{New J. Phys.} \textbf{\bibinfo{volume}{10}},
  \bibinfo{pages}{023011} (\bibinfo{year}{2008}).

\bibitem[{\citenamefont{Buscemi}(2006)}]{Buscemi:06}
\bibinfo{author}{\bibfnamefont{F.}~\bibnamefont{Buscemi}},
  \bibinfo{journal}{Phys. Lett. A} \textbf{\bibinfo{volume}{360}},
  \bibinfo{pages}{256} (\bibinfo{year}{2006}).

\bibitem[{\citenamefont{Laflamme et~al.}(1996)\citenamefont{Laflamme, Miquel,
  Paz, and Zurek}}]{Laflamme:96}
\bibinfo{author}{\bibfnamefont{R.}~\bibnamefont{Laflamme}},
  \bibinfo{author}{\bibfnamefont{C.}~\bibnamefont{Miquel}},
  \bibinfo{author}{\bibfnamefont{J.}~\bibnamefont{Paz}}, \bibnamefont{and}
  \bibinfo{author}{\bibfnamefont{W.}~\bibnamefont{Zurek}},
  \bibinfo{journal}{Phys. Rev. Lett.} \textbf{\bibinfo{volume}{77}},
  \bibinfo{pages}{198} (\bibinfo{year}{1996}).

\bibitem[{\citenamefont{Bernstein}(2005)}]{Bernstein:book}
\bibinfo{author}{\bibfnamefont{D.~S.} \bibnamefont{Bernstein}},
  \emph{\bibinfo{title}{Matrix Mathematics}} (\bibinfo{publisher}{Princeton
  University Press}, \bibinfo{address}{Princeton}, \bibinfo{year}{2005}).

\end{thebibliography}

\end{document}